\documentclass[preprint]{aastex}
\pdfoutput=1 
\usepackage[usenames,dvipsnames]{color}
\usepackage{natbib,graphicx,latexsym,lscape,pdflscape,url}
\usepackage[breaklinks,colorlinks,urlcolor=blue,citecolor=black,linkcolor=blue]{hyperref}

\newcommand{\WISE}{{\sl WISE}}
\newcommand{\Gaia}{{\sl Gaia}}

\newcommand{\Msun}{\mbox{$M_{\sun}$}}

\newcommand{\Lsun}{\mbox{$L_{\sun}$}}

\newcommand{\Mjup}{\mbox{$M_{\rm Jup}$}}
\newcommand{\Rjup}{\mbox{$R_{\rm Jup}$}}
\newcommand{\degree}{\mbox{$^{\circ}$}}

\newcommand{\kms}{\mbox{km\,s$^{-1}$}}

\newcommand{\masyr}{\hbox{mas\,yr$^{-1}$}}

\newcommand{\Mtot}{\mbox{$M_{\rm tot}$}}

\newcommand{\fbol}{\mbox{$f_{\rm bol}$}}
\newcommand{\Lbol}{\mbox{$L_{\rm bol}$}}

\newcommand{\Mbol}{\mbox{$M_{\rm bol}$}}
\newcommand{\Teff}{\mbox{$T_{\rm eff}$}}
\newcommand{\logg}{\mbox{$\log{g}$}}

\newcommand{\vlg}{\textsc{vl-g}}
\newcommand{\intg}{\textsc{int-g}}
\newcommand{\fldg}{\textsc{fld-g}}

\newcommand{\obj}{LSPM~J1314+1320}

\shorttitle{Dynamical Masses of \obj{AB}}
\shortauthors{Dupuy et al.}

\begin{document}

\title{High-Precision Radio and Infrared Astrometry of LSPM~J1314+1320AB -- II: 
Testing Pre--Main-Sequence Models at the Lithium Depletion Boundary with Dynamical Masses\altaffilmark{*}}

\author{Trent J.\ Dupuy,\altaffilmark{1}
        Jan Forbrich,\altaffilmark{2,3}
        Aaron Rizzuto,\altaffilmark{1}
        Andrew W.\ Mann,\altaffilmark{1}
        Kimberly Aller,\altaffilmark{4}
        Michael C.\ Liu,\altaffilmark{4}
        Adam L.\ Kraus,\altaffilmark{1} and
        Edo Berger\altaffilmark{3}}

      \altaffiltext{*}{Data presented herein were obtained at the
        W.~M.\ Keck Observatory, which is operated as a scientific
        partnership among the California Institute of Technology, the
        University of California, and the National Aeronautics and
        Space Administration. The Observatory was made possible by the
        generous financial support of the W.~M.\ Keck Foundation.}

      \altaffiltext{1}{The University of Texas at Austin, Department
        of Astronomy, 2515 Speedway C1400, Austin, TX 78712, USA}

      \altaffiltext{2}{University of Vienna, Department of
        Astrophysics, T\"{u}rkenschanzstr. 17, 1180 Vienna, Austria}

      \altaffiltext{3}{Harvard-Smithsonian Center for Astrophysics, 60
        Garden Street, Cambridge, MA 02138, USA}

      \altaffiltext{4}{Institute for Astronomy, University of Hawai`i,  
        2680 Woodlawn Drive, Honolulu, HI 96822, USA}

\begin{abstract}

  We present novel tests of pre--main-sequence models based on
  individual dynamical masses for the M7 binary
  LSPM~J1314+1320AB. Joint analysis of our Keck adaptive optics
  astrometric monitoring along with Very Long Baseline Array radio
  data from a companion paper yield component masses of
  $92.8\pm0.6$\,\Mjup\ ($0.0885\pm0.0006$\,\Msun) and
  $91.7\pm1.0$\,\Mjup\ ($0.0875\pm0.0010$\,\Msun) and a parallactic
  distance of $17.249\pm0.013$\,pc. We also derive component
  luminosities that are consistent with the system being coeval at an
  age of $80.8\pm2.5$\,Myr, according to BHAC15 evolutionary
  models. The presence of lithium is consistent with model
  predictions, marking the first time the theoretical lithium
  depletion boundary has been tested with ultracool dwarfs of known
  mass. However, we find that the average evolutionary model-derived
  effective temperature ($2950\pm5$\,K) is 180\,K hotter than we
  derive from a spectral type--\Teff\ relation based on BT-Settl
  models ($2770\pm100$\,K). We suggest that the dominant source of
  this discrepancy is model radii being too small by $\approx$13\%. In
  a test that mimics the typical application of evolutionary models by
  observers, we derive masses on the H-R diagram using the luminosity
  and BT-Settl temperature. The estimated masses are
  $46^{+16}_{-19}$\% (2.0$\sigma$) lower than we measure dynamically
  and would imply that this is a system of $\approx$50\,\Mjup\ brown
  dwarfs, highlighting the large systematic errors possible when
  inferring masses from the H-R diagram. This is first time masses
  have been measured for ultracool ($\geq$M6) dwarfs displaying
  spectral signatures of low gravity. Based on features in the
  infrared, LSPM~J1314+1320AB appears higher gravity than typical
  Pleiades and AB~Dor members, opposite the expectation given its
  younger age. The components of LSPM~J1314+1320AB are now the
  nearest, lowest mass pre--main-sequence stars with direct mass
  measurements.

\end{abstract}



\section{Introduction}

A major goal of stellar astrophysics is to understand the early
evolution of stars, before they reach a stable equilibrium on the main
sequence.  In theory, the fundamental parameters of mass, composition,
and angular momentum uniquely determine the course of all stellar
evolution, including the pre--main-sequence phase.  Binary stars are
perhaps the most useful empirical calibrators available for testing
stellar models, as their components share a common age and composition
and dynamical masses can be derived from their orbital motion.  While
dozens of mass measurements have been obtained for pre--main-sequence
stars \citep[e.g., see reviews from][]{2004ApJ...604..741H,
  2007prpl.conf..411M, 2012MNRAS.420..986G, 2014NewAR..60....1S}, most
of these are for stars more massive than 0.5\,\Msun.  Steady progress
has been made to push measurements to lower masses
\citep[e.g.,][]{2000ApJ...545.1034S, 2006Natur.440..311S,
  2015ApJ...807....3K, 2015A&A...584A.128L, 2016ApJ...816...21D}.
However, there are still only a handful of masses measured for stars
at or below the 0.2--0.3\,\Msun\ peak in the initial mass function
\citep{2010ARA&A..48..339B}, leaving pre--main-sequence models for a
large fraction of stars poorly constrained.

Previous work to measure the masses of pre--main-sequence stars has
mostly focused on star forming regions, like the nearby Taurus--Auriga
and Scorpius--Centaurus--Lupus--Crux complexes with ages of
$\sim$1--10\,Myr and distances of $\sim$150\,pc
\citep[e.g.,][]{2008AJ....135.1659S, 2016ApJ...817..164R,
  2016ApJ...818..156C}.  Over the last two decades a growing number of
young stars much closer to the Sun and with wider ranging ages
($\sim$8--150\,Myr) have been identified
\citep[e.g.,][]{2004ARA&A..42..685Z, 2008hsf2.book..757T}.  The
proximity of these stars offers many benefits, including the
possibility of spatially resolving binaries with smaller semimajor
axes and correspondingly shorter orbital periods for dynamical mass
determinations.

\obj\ was first identified as a star exhibiting proper motion
$>$200\,\masyr\ by \citet{1979NLTT..C02....0L}.
\citet{2006MNRAS.368.1917L} included it as a candidate late-type star
in their high angular resolution survey using lucky imaging at
$i^{\prime}$ and $z^{\prime}$ bands, where their sample was selected
from objects with red $V-K$ colors in the LSPM catalog
\citep{2005AJ....129.1483L}.  \citet{2006MNRAS.368.1917L} estimated a
spectral type of M6 for \obj\ from its $V-K$ color and discovered that
it was a binary with a separation of 130\,mas and flux ratios of
$\Delta{i^{\prime}} = 0.93\pm0.25$\,mag and $\Delta{z^{\prime}} =
0.97\pm0.25$\,mag.  Meanwhile, \citet{2005AJ....130.1680L} had
identified \obj\ as a potential nearby star with an estimated distance
of 9.7\,pc, leading them to obtain spectra and astrometry that
revealed a spectral type of M7, H$\alpha$ in emission, and a
parallactic distance of $16.4\pm0.8$\,pc
\citep{2009AJ....137.4109L}. They concluded that their original
photometric distance estimate was much smaller than the parallactic
distance due to unresolved binarity and/or extreme youth.

The first definitive evidence for the youth of \obj{AB} came from
analysis by \citet{2014ApJ...783...27S} of its published optical
spectra and their own near-infrared (NIR) spectrum, all spatially
unresolved.  These spectra display weak alkali lines, which are
indicative of low surface gravity, and strong \ion{Li}{1} absorption.
Although \citet{2012AJ....143...80S} had previously identified
\obj{AB} as a likely new member of the AB Doradus moving group
($\sim$150\,Myr), subsequent analysis and updated proper motion data
has led to the conclusion that it cannot be confidently associated
with any known group \citep{2014ApJ...783..121G, 2014ApJ...783...27S}.
\citet{2011ApJ...741...27M} has also identified \obj{AB} as a source
of bright ($\sim$1\,mJy), persistent radio emission with a flat
spectrum across a wide range of frequencies (1.43--22.5\,GHz),
suggesting high levels of magnetic activity and a stable, large-scale
magnetosphere \citep{2014ApJ...785....9W, 2015ApJ...799..192W}.
Therefore, \obj{AB} is potentially useful as a benchmark not only for
evolutionary models but also for models of stellar magnetism.

We present here spatially resolved relative astrometric monitoring of
\obj{AB} obtained with Keck AO that allows us to determine its orbit
and thereby its dynamical total mass (\Mtot).  We combine our Keck
data with spatially resolved Very Long Baseline Array (VLBA) absolute
astrometry from a companion paper (Forbrich et al.\ 2016, submitted;
hereinafter Paper~I) to simultaneously model the orbital motion,
proper motion, and parallax and thereby derive individual dynamical
masses of $\lesssim$1\% precision.  We confirm that \obj{AB} is
unambiguously in the pre--main-sequence phase of its evolution and use
our dynamical masses to perform tests of models, examining properties
such as luminosity, temperature, and lithium depletion.


\section{Observations \label{sec:obs}}

\subsection{Keck/NIRC2 Astrometry \label{sec:keck}}

\begin{figure}
\centerline{
  \includegraphics[height=1.5in,angle=0]{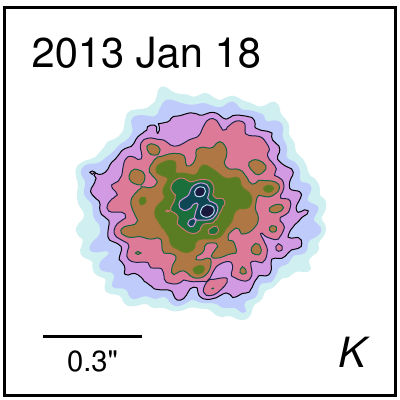}  \hskip 0.04in
  \includegraphics[height=1.5in,angle=0]{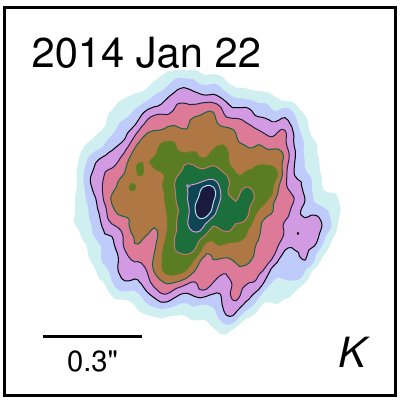}  \hskip 0.04in
  \includegraphics[height=1.5in,angle=0]{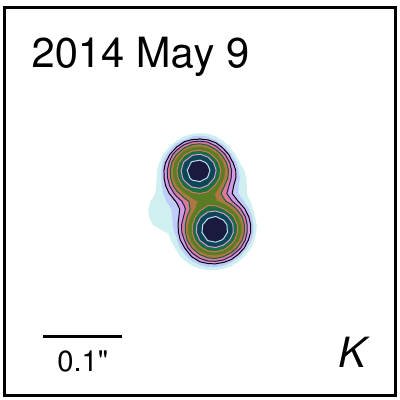}   \hskip 0.04in
  \includegraphics[height=1.5in,angle=0]{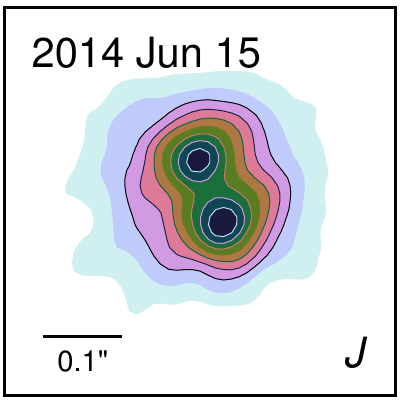}  \hskip 0.04in
}
\vskip 0.05in
\centerline{
  \includegraphics[height=1.5in,angle=0]{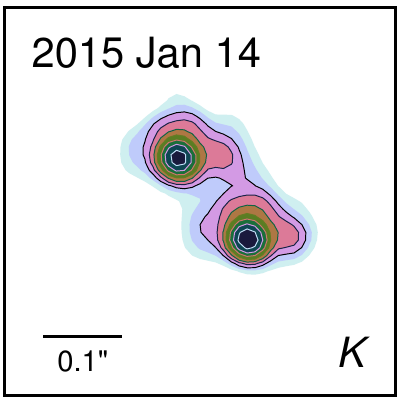}  \hskip 0.04in
  \includegraphics[height=1.5in,angle=0]{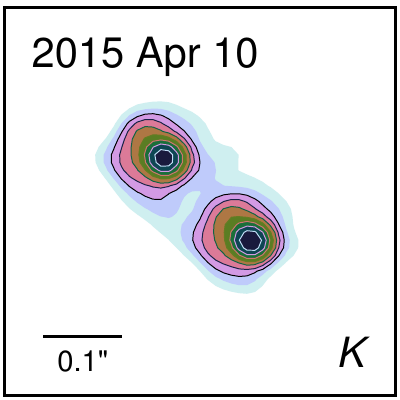}  \hskip 0.04in
  \includegraphics[height=1.5in,angle=0]{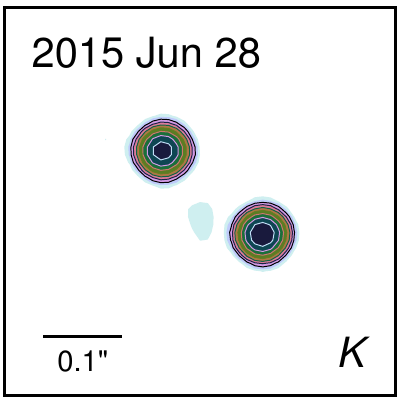}  \hskip 0.04in
}
\vskip 0.3in
\caption{\normalsize Contour plots of our Keck AO interferograms
  (first two panels) and images used to derive relative astrometry and
  flux ratios (Table~\ref{tbl:keck}).  Contours are in logarithmic
  intervals from unity to 10\% of the peak flux in each band.  The
  image cutouts are all 0\farcs5 across and interferogram cutouts are
  1$\farcs$2 across.  In the interferograms, the binary can be seen by
  eye as an elongation or double peak in the center of the
  point-spread function. All cutouts have the same native pixel scale,
  and here we have rotated them so that north is up. \label{fig:keck}}

\end{figure}

We have been monitoring the resolved orbital motion of both components
of \obj{AB} using Keck adaptive optics (AO) with the facility NIR
camera NIRC2.  We used both direct imaging and non-redundant aperture
masking to measure the binary's separation, position angle (PA), and
flux ratio in $J$ and $K$ bands.  Typical examples of our images and
masking interferograms from each epoch are shown in
Figure~\ref{fig:keck}.  Analysis of masking data was done using a
pipeline similar to that in previous papers containing NIRC2 masking
data \citep[e.g.,][]{2008ApJ...678..463I, 2008ApJ...678L..59I} and is
described in detail in Section~2.2 of \citet{2009ApJ...699..168D}.
For imaging data, we used the same methods described in our previous
work \citep[e.g., see][]{2009ApJ...692..729D, 2016ApJ...817...80D}.
Briefly, after a standard reduction of the images (dark subtraction
and flat fielding) we fit a three-component two-dimensional Gaussian
model to the two binary components.  For the last three epochs, when
the binary separation was widest, we were able to use the StarFinder
package \citep{2000A&AS..147..335D} instead.  StarFinder iteratively
solves for both the binary parameters and an image of the point-spread
function.  We then corrected our derived $(x,y)$ positions using the
NIRC2 distortion solution of \citet{2010ApJ...725..331Y}, which has a
pixel scale of $9.952\pm0.002$\,mas\,pixel$^{-1}$ and a
$+0\fdg252\pm0\fdg0.009$ offset added to the orientation given in the
NIRC2 image headers.  The last epoch was obtained after a major
realignment of the Keck AO system on 2015~Apr~13.  Therefore, for this
epoch, we used an updated distortion solution from Service et
al.\ (2016, submitted), which has a pixel scale of
$9.971\pm0.004$\,mas\,pixel$^{-1}$ and orientation of
$+0\fdg262\pm0\fdg0.020$.

Table~\ref{tbl:keck} summarizes the resolved astrometry and flux
ratios derived from our Keck observations. As an estimate of our
uncertainties, we used the rms of the best-fit binary parameters at a
given epoch across the multiple images or masking interferograms. To
vet these errors, we performed a standard 7-parameter orbit
least-squares fit \citep[see Section~3.1 of][]{2010ApJ...721.1725D} to
our Keck astrometry. We found a value of $\chi^2 = 3.5$ for 7~degrees
of freedom (dof), which has a probability of $p(\chi^2) = 0.83$, so we
concluded that our Keck astrometric errors are
reasonable.\footnote{When we quote $p(\chi^2)$, it is the probability
  of obtaining a value for $\chi^2$ as high or higher than the
  observed value given the degrees of freedom.}

Our $K$-band flux ratios are somewhat inconsistent with being constant
across all epochs, with $\chi^2 = 12.4$ (6~dof). To achieve $p(\chi^2)
= 0.5$, a systematic error of 0.011\,mag added in quadrature is
required. Therefore, in the following analysis we use $\Delta{K} =
0.080\pm0.022$\,mag, which is the weighted average after adding this
error floor in quadrature to the individual values. We conservatively
use the rms of the measurements as the uncertainty since it
encompasses both the fitting errors and the potential for variability.
Finally, we note that our infrared flux ratios seem to be at odds with
the optical flux ratios of $\Delta{i^{\prime}} = 0.93\pm0.25$\,mag and
$\Delta{z^{\prime}} = 0.97\pm0.25$\,mag from
\citet{2006MNRAS.368.1917L}.  However, for another binary in their
survey, LHS~1901AB, they also report very unequal flux ratios
\citep[$\Delta{i^{\prime}} = 1.3\pm0.7$\,mag and $\Delta{z^{\prime}} =
1.3\pm0.7$\,mag;][]{2008MNRAS.384..150L} while AO imaging gives flux
ratios in $JHK$ of $\approx$0.1\,mag \citep{2006A&A...460L..19M,
  2010ApJ...721.1725D}.  This suggests that the apparent discrepancy
may simply be due to systematic errors in the flux ratios derived from
Lucky imaging, so we do not use them in our analysis.

\subsection{UH~2.2-m/SNIFS \& IRTF/SpeX Spectroscopy \label{sec:spectra}}

We obtained optical and near-infrared (NIR) spectra of \obj{AB} as
part of larger follow-up program of nearby bright M~dwarfs from
\citet{2011AJ....142..138L}. In the optical we used the SuperNova
Integral Field Spectrograph \citep[SNIFS;][]{2002SPIE.4836...61A,
  2004SPIE.5249..146L} on the University of Hawaii 2.2-m telescope on
Maunakea, Hawaii, on 2015~Jan~9~UT. SNIFS provides simultaneous
coverage from 3200--9700\,\AA\ at a resolution of
$R\simeq1000$. Details of our SNIFS observations and reduction can be
found in \citet{2001MNRAS.326...23B} and \citet{2014MNRAS.443.2561G},
which we briefly summarize here. The pipeline detailed in
\citet{2001MNRAS.326...23B} performs dark, bias, and flat-field
corrections, cleaned the data of bad pixels and cosmic rays, then fits
and extracts the integral field unit spaxels into a one-dimensional
spectrum. The \citet{2014MNRAS.443.2561G} reduction takes this
spectrum and performs flux calibration and telluric correction based
on white dwarf standards taken throughout the night and a model of the
atmosphere above Maunakea \citep{2013A&A...549A...8B}. The final
reduced spectrum of \obj{AB} has a signal-to-noise ratio (SNR) of
$>$100 per pixel redward of 6000\,\AA.

We obtained a NIR spectrum of \obj{AB} using the SpeX spectrograph
\citep{2003PASP..115..362R} at the NASA Infrared Telescope Facility
(IRTF) on Maunakea, Hawaii, on 2013~May~16~UT. Our observations were
taken in the short cross-dispersed (SXD) mode using the
$0.3\arcsec\times15\arcsec$ slit ($R\simeq2000$), yielding
simultaneous coverage from 0.8--2.4\,\micron, with a small gap near
1.8\,\micron\ due to non-overlapping SXD orders. The target was placed
at two positions along the slit (A and B) and observed in an ABBA
pattern in order to subsequently subtract the sky background. Six
exposures were taken this way, yielding a SNR $>200$ per pixel in the
$H$ and $K$ bands. To correct for telluric lines, we observed an A0V
star immediately after the target.

Our SpeX spectrum was extracted using the SpeXTool package
\citep{2004PASP..116..362C}, which performs flat-field correction,
wavelength calibration, sky subtraction, and extraction of the
one-dimensional spectrum.  Multiple exposures were combined using the
\texttt{xcombspec} routine. A telluric correction spectrum was
constructed from the A0V star and applied to the target spectrum using
the \texttt{xtellcor} package \citep{2003PASP..115..389V}. Separate
orders were stacked using the \texttt{xcombspec} tool, which also
shifts the flux level in different orders to match each other. These
corrections were 1\% or less per order.


\section{Results}

\subsection{MCMC Astrometric Orbit \& Parallax Analysis \label{sec:orbit}}

We combined our Keck/NIRC2 relative astrometry of \obj{AB} with VLBA
absolute astrometry from Paper~I to perform a joint analysis of the
orbit and parallax of the system. The VLBA observations only detect
one component of the binary at all epochs, and we identify it as the
secondary component \obj{B} (defined as the component that is fainter
in our $J$- and $K$-band Keck data). This is because the orbital
motion seen in the VLBA data is to the northeast over 2013--2014, and
our Keck astrometry over the same time period indicates that the
secondary was moving to the northeast relative to the primary. In
addition to our Keck astrometry here, we also used results from
\citet{2006MNRAS.368.1917L} who measured a binary separation of
$130\pm20$\,mas and PA of $46\fdg0\pm2\fdg0$ from their Lucky imaging
data on 2005~Jun~15~UT.  The combined data set contains a total of
7~epochs of resolved astrometry spanning 10.03\,yr and 9~epochs of
absolute astrometry spanning 4.65\,yr.

Our astrometric model includes 15 total parameters, and six of these
are orbit parameters shared between the Keck and VLBA data. The six
common parameters are orbital period ($P$), eccentricity ($e$),
inclination ($i$), argument of periastron ($\omega$), mean longitude
at the reference epoch ($\lambda_{\rm ref}$), and PA of the ascending
node ($\Omega$). We defined the reference epoch for our model as
$t_{\rm ref} = 2455197.5$\,JD (i.e., calendar date 2010.0). The
semimajor axes of the primary and secondary components about their
barycenter are denoted $a_1$ and $a_2$, respectively. Since our Keck
data measure relative orbital motion, the Keck semimajor axis
parameter was $a = a_1 + a_2$, while the VLBA semimajor axis parameter
was simply $a_2$. The remaining parameters are only used to model the
VLBA data: RA and Dec at the reference epoch ($\alpha_{\rm 2010}$,
$\delta_{\rm 2010}$); proper motion in RA and Dec
($\mu_{\alpha\cos\delta}$, $\mu_{\delta}$); parallax ($\pi$); and
systematic error parameters in RA and Dec ($\sigma_{\alpha}$,
$\sigma_{\delta}$). These VLBA error parameters are needed to model
systematic errors in the absolute astrometric calibration, which are
expected to be comparable from epoch to epoch. In order to properly
mitigate our model from preferring extremely large values of these
error parameters, we penalized the logarithm of the likelihood by
$\log(1/\sigma^2)$, where $\sigma$ is the quadrature sum of both error
parameters. The parallax factor at each epoch was calculated from the
JPL ephemeris DE405 as described in Equations~1 and 2 of
\citet{2012ApJS..201...19D}.

\begin{figure} 

  \centerline{\includegraphics[width=6.5in,angle=0]{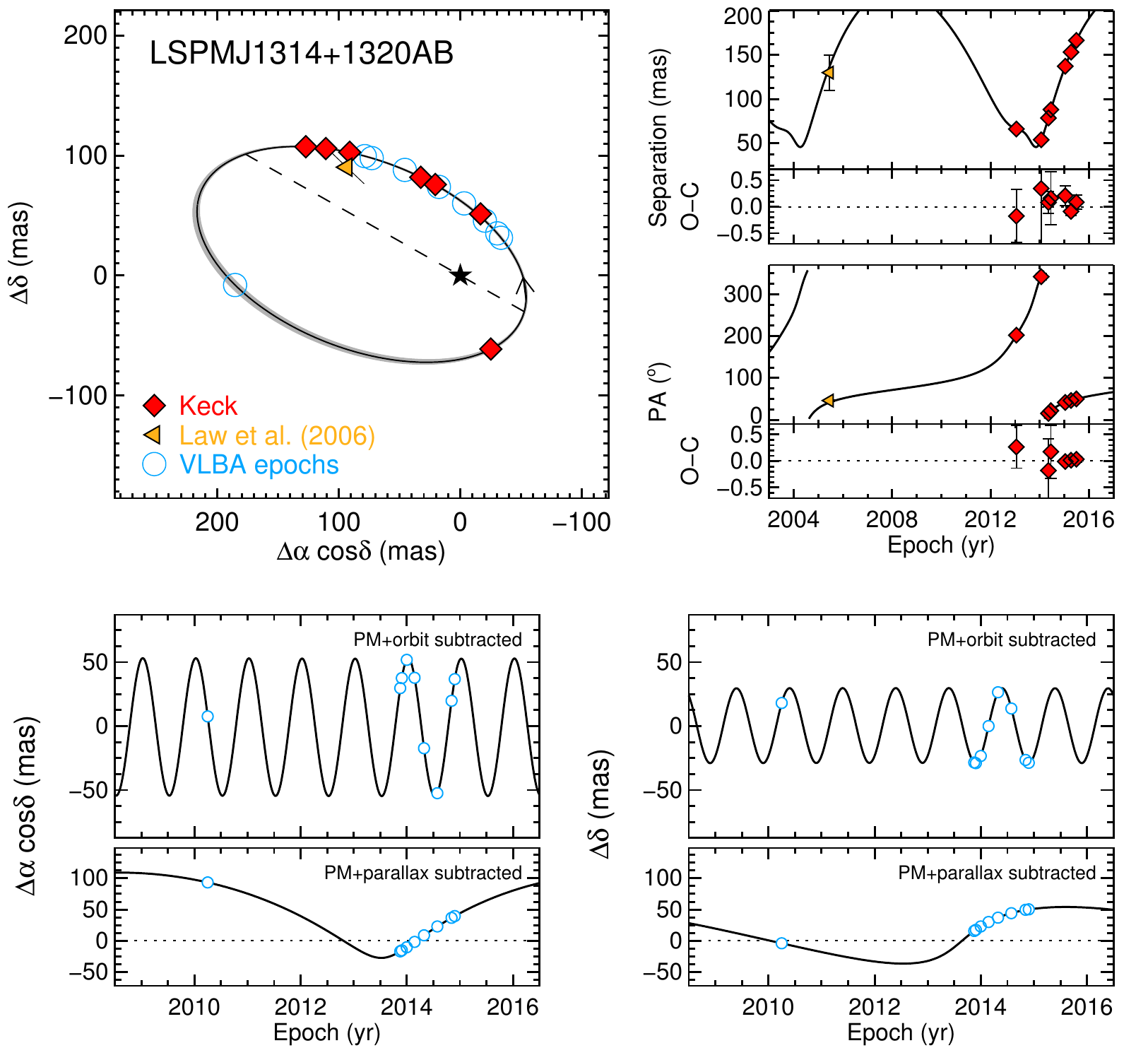}}

  \caption{\normalsize \emph{Top left:} resolved relative astrometry
    (filled symbols) shown alongside the best fit orbit (thick black
    line) and 100 randomly drawn orbits from our MCMC chain (thin gray
    lines).  The plotting symbols typically are larger than the error
    bars.  Open blue circles indicate the epochs of VLBA astrometry of
    the secondary from Paper~I.  \emph{Top right:}~our relative
    astrometry shown as a function of time (top sub-panels) and after
    subtracting the best-fit orbit solution (bottom sub-panels).
    \emph{Bottom:} VLBA astrometry of the secondary from Paper~I.  Top
    panels show the data with the proper motion and orbital motion
    subtracted in order to display the best-fit parallax solution
    (thick black line).  Bottom panels show the orbit of the secondary
    component after subtracting the best-fit parallax and proper
    motion.  \label{fig:orbit}}

\end{figure}

As in our previous work on joint analysis of relative and absolute
astrometry \citep{2015ApJ...805...56D}, we used the affine invariant
ensemble sampler \texttt{emcee~v2.1.0} \citep{2013PASP..125..306F} to
perform our Markov Chain Monte Carlo (MCMC) analysis. We used $10^3$
walkers (chains) with $10^5$ steps each, saving every 500th step for
use in our analysis and removing the first 10\% of all walkers as the
burn-in portion. Table~\ref{tbl:mcmc} summarizes the resulting
posterior distributions and the priors on all of our parameters. We
report both the best-fit parameters and the credible intervals that
encompass 68.3\% and 95.4\% of each parameter's posterior values.  The
best-fit parameters give a total $\chi^2 = 17.8$ for 19 dof,
$p(\chi^2) = 0.53$, including both the Keck data and the VLBA data
with the systematic errors added in quadrature to the nominal VLBA
measurement errors.
Figure~\ref{fig:orbit} displays this best-fit orbit and parallax
solution alongside both the Keck and VLBA astrometry.

Table~\ref{tbl:mcmc} also gives credible intervals and best-fit values
for a number of additional parameters that can be derived directly
from our fitted parameters, e.g., the distance computed from the
absolute VLBA parallax ($d=1/\pi$). Most notably, we compute a precise
total system mass from the Keck orbit and VLBA parallax, $\Mtot =
(a/\pi)^3 P^{-2}$, and individual masses for the components by
combining the VLBA and Keck orbits ($M_1 = \frac{a_2}{a}\Mtot$, $M_2 =
\frac{a-a_2}{a}\Mtot$). Because of the very high precision of the VLBA
parallax ($\sigma_{\pi}/\pi = 7.8\times10^{-4}$), the uncertainties in
these masses are dominated by the uncertainty in the orbital period
($\sigma_P/P = 7.7\times10^{-3}$), total semimajor axis ($\sigma_a/a =
3.5\times10^{-3}$), and secondary semimajor axis ($\sigma_{a_2}/a_2 =
4.5\times10^{-3}$). Accounting for covariances between these
parameters, the derived primary mass is the most precise
($\sigma_{M_1}/M_1 = 6.8\times10^{-3}$), followed by the total mass
($\sigma_{\Mtot}/\Mtot = 8.7\times10^{-3}$) and secondary mass
($\sigma_{M_2}/M_2 = 1.1\times10^{-2}$).
The mass ratio ($q \equiv M_2/M_1 = \frac{a}{a_2} - 1 =
0.989\pm0.007$) is consistent with the radio-emitting secondary that
is fainter in the NIR being the less massive component, and 95.4\% of
our MCMC posterior values having $q<1$.
The derived primary and secondary masses are $92.8\pm0.6$\,\Mjup\
($0.0885\pm0.0006$\,\Msun) and $91.7\pm1.0$\,\Mjup\
($0.0875\pm0.0010$\,\Msun), respectively. These individual masses are
very nearly equal, within 2\% of each other at 90\% confidence. For
additional discussion of the parallax and proper motion determined
from this joint analysis see Paper~I.

\subsection{Other Empirically Determined Properties \label{sec:lbol}}

Our new AO imaging allows us to compute resolved photometry for the
two components of \obj{AB}. Our $J$- and $K$-band flux ratios
($\Delta{J} = 0.08\pm0.04$\,mag, $\Delta{K} = 0.071\pm0.013$\,mag) are
consistent with and more precise than the values of $\Delta{J} =
0.10\pm0.11$\,mag and $\Delta{K_S} = 0.10\pm0.21$\,mag from
\citet{2014ApJ...783...27S}. Using our $\Delta{J}$ and $\Delta{K}$
values and their $H$-band flux ratio, $\Delta{H} = 0.03\pm0.06$\,mag,
we computed resolved $JHK$ photometry for the components of \obj{AB}.
For the integrated-light photometry we used the 2MASS Point Source
Catalog \citep{2003tmc..book.....C} along with 2MASS-to-MKO
photometric system conversions calculated from our SpeX spectrum. We
list the resulting resolved and integrated-light photometry on both
the MKO and 2MASS photometric systems in Table~\ref{tbl:props}.

\begin{figure} 

  \centerline{\includegraphics[width=6.5in]{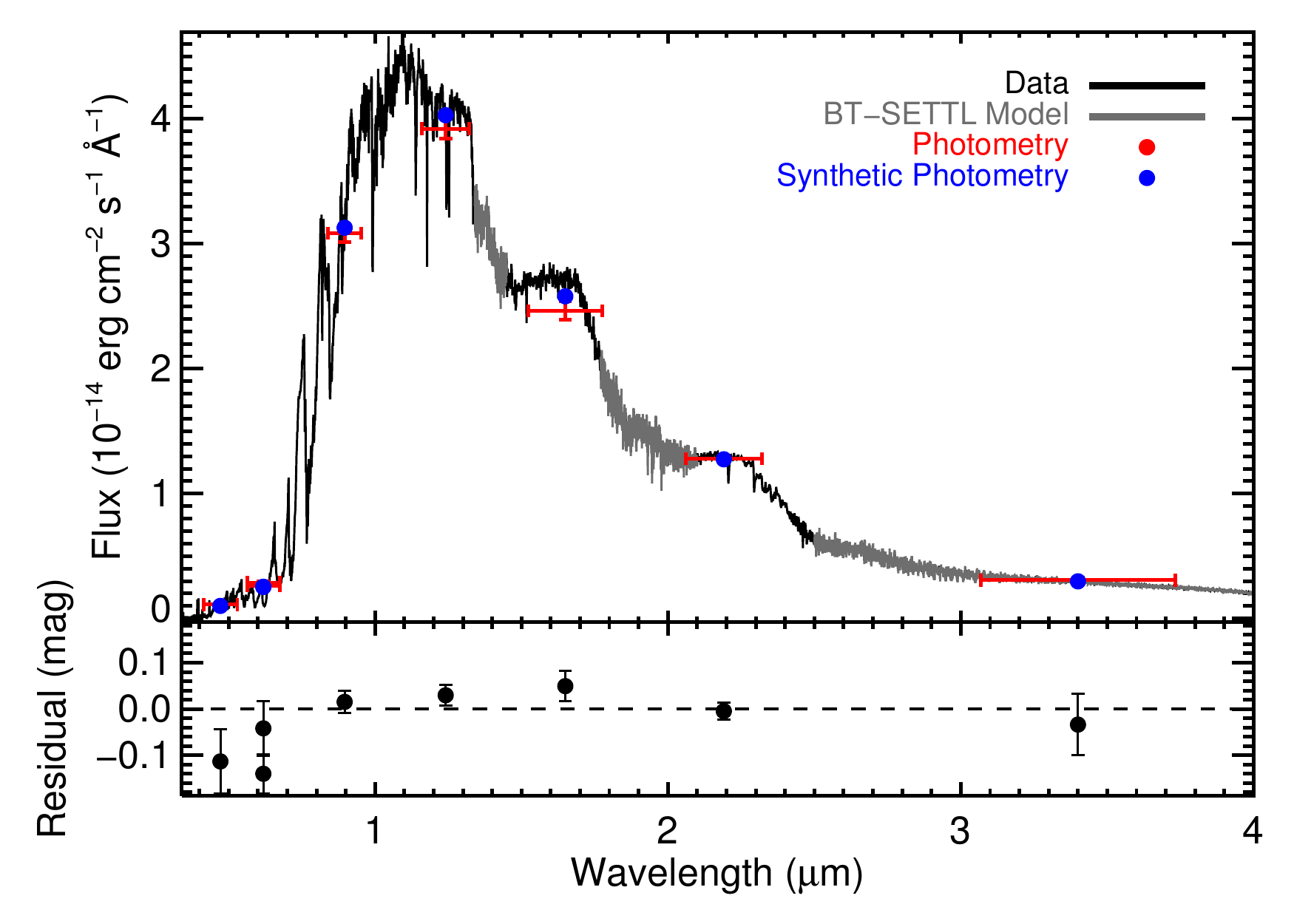}}

  \caption{\normalsize Flux-calibrated spectrum of \obj{AB} from which
    we determine its integrated-light bolometric flux.  Combined
    SNIFS+SpeX data are shown in black, and spectral regions replaced
    by models are shown in gray. Literature photometry is shown in
    red, with the horizontal bars indicating the width of the filter
    and vertical error bars representing combined measurement and zero
    point errors.  Blue points indicate the corresponding synthetic
    fluxes derived from the spectrum. The bottom panel shows the
    photometry residuals.  This plot is truncated at 4.0\,\micron\ for
    display purposes, but we include all \WISE\ photometry in our
    analysis. \label{fig:lbol}}

\end{figure}

In order to compute the bolometric flux of \obj{AB} in integrated
light, we combined and absolutely calibrated our optical and NIR
spectra following the method outlined in \citet{2015ApJ...804...64M}.
Briefly, we first collected published photometry from 2MASS, SDSS
\citep{2012ApJS..203...21A}, the AAVSO All-Sky Photometric Survey
\citep[APASS;][]{2012JAVSO..40..430H}, and the \textsl{Wide-field
  Infrared Survey Explorer} \citep[\WISE;][]{2010AJ....140.1868W}.  We
converted this photometry to fluxes using the relevant zero-points,
and we also calculated synthetic fluxes from our spectra using the
corresponding filter profiles \citep{2003AJ....126.1090C,
  2011ApJ...735..112J, 2015PASP..127..102M}.
At this point we noticed that the SDSS $i$-band measurement was highly
discrepant ($\approx$3--4\,mag) with our spectrum and with the
surrounding photometry, so we excluded it from our analysis.
Taking account of both random and correlated errors (flux calibration
and optical variability) we scaled the optical and NIR spectra to
match the photometry and to match each other in the region overlapping
between SNIFS and SpeX (0.80--0.95\,\micron).  We replaced regions of
high telluric contamination and those not covered by our spectra with
a best-fit atmospheric model from the BT-Settl grid
\citep{2011ASPC..448...91A, 2012RSPTA.370.2765A}, where the best-fit
here was $\Teff = 2800$\,K and $\logg = 4.5$\,dex.  The final
calibrated and combined spectrum is shown in Figure~\ref{fig:lbol}.
We calculated the bolometric flux (\fbol) by integrating over the
combined spectrum, accounting for errors in the flux calibration,
optical variability of 1.5\% \citep{2011ApJ...741...27M,
  2015ApJ...799..192W}, and errors introduced by replacing regions of
spectra with a BT-Settl model. This yielded $\fbol = (5.12\pm0.10)
\times 10^{-10}$\,erg\,s$^{-1}$\,cm$^{-2}$.

We apportioned this integrated-light bolometric flux to the individual
components by deriving a bolometric flux ratio from our $K$-band flux
ratio.  We examined BT-Settl atmosphere models with $\Teff =
2700$--2900\,K and $\logg = 4.5$--5.0\,dex.  Relative to the best-fit
$\Teff = 2800$\,K and $\logg = 4.5$\,dex model, the other five models
had comparable or slightly larger amplitude $K$-band flux ratios
spanning $\Delta{K} = -0.10$\,mag to 0.12\,mag for hotter and cooler
models, respectively.  Among these models, the relationship between
their bolometric magnitude differences and $K$-band magnitude
differences was consistently $\Delta{\Mbol} - \Delta{K} =
-0.042\pm0.011$\,mag.  In comparison to our integrated-light
bolometric magnitude error of 0.025\,mag, the correction itself is
small and the uncertainty in the correction is essentially negligible
(though we do account for it in our analysis).  Combining this
correction factor with our $K$-band photometry, we find
$\Delta{\log(\Lbol)} = -0.015\pm0.010$\,dex, with the resultant
individual luminosities given in Table~\ref{tbl:props}.

\subsection{Spectral Type \& Gravity Classification \label{sec:spt}}

\begin{figure} 

  \centerline{\includegraphics[width=5.0in]{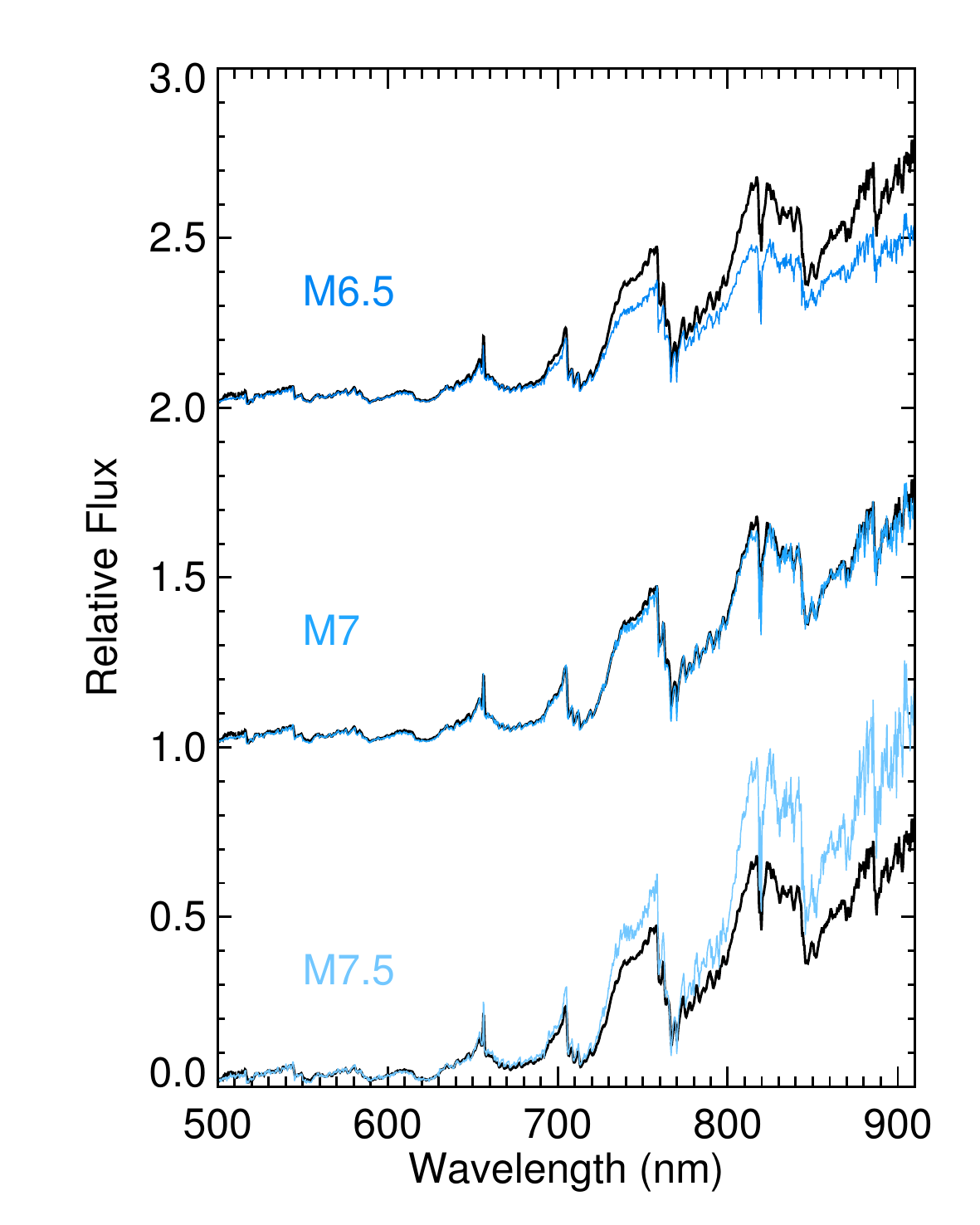}}

  \caption{\normalsize SNIFS integrated-light spectrum of \obj{AB}
    shown in black compared to the M6.5, M7, and M7.5 spectral
    standards from \citet{2007AJ....133..531B} shown in shades of
    blue.  Interpolating these standards in a least-squares fit we
    determined a spectral type of M$7.0\pm0.2$. \label{fig:spt}}

\end{figure}

We determined the integrated-light spectral type of \obj{AB} from our
optical and NIR data.  We compared our optical SNIFS spectrum to the
M~dwarf spectral library of \citet{2007AJ....133..531B}.  To find a
best fit spectral type and uncertainty we used the least-squares
package MPFIT in IDL \citep{2009ASPC..411..251M}.  In performing this
fit we allowed the numerical spectral type to vary continuously,
interpolating between standards to create the comparison spectra.  We
excluded wavelengths near H$\alpha$, allowed for a small wavelength
shift to account for radial velocity, and neglect extinction given the
small distance.  We found a best fit spectral type for \obj{AB} of
M$7.0\pm0.2$, and Figure~\ref{fig:spt} shows our data in comparison to
the best-fit and adjacent spectral standards.

\begin{figure} 

  \centerline{\includegraphics[width=6.0in]{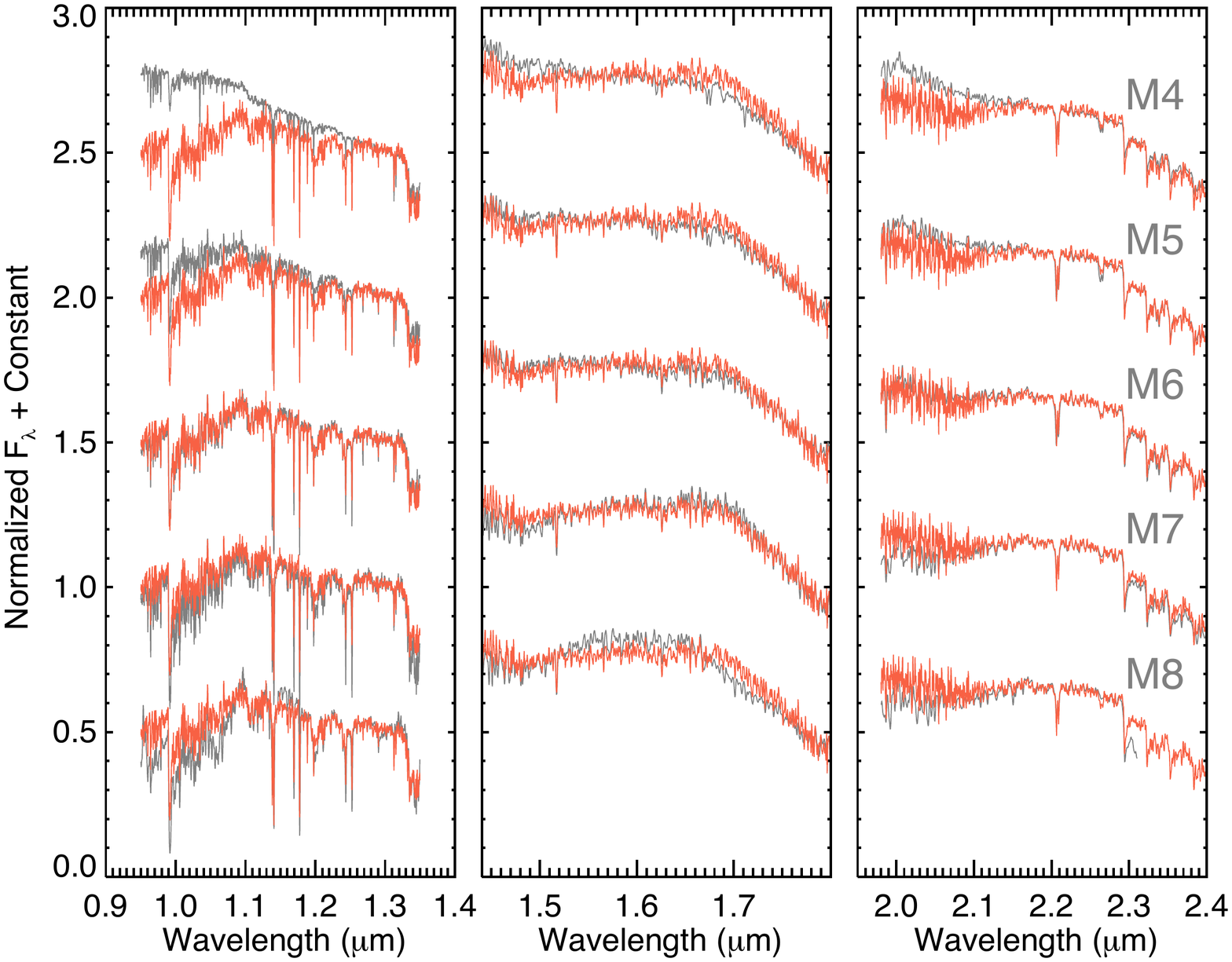}}
  \vskip 0.2in

  \caption{\normalsize Integrated-light SpeX spectrum of \obj{AB}
    (orange) compared to infrared spectral standards from
    \citet[][gray]{2013ApJ...772...79A}.  The spectral type on this
    system is M$6\pm1$. \label{fig:ir-spt}}

\end{figure}

\begin{figure} 

  \centerline{\includegraphics[height=3.5in]{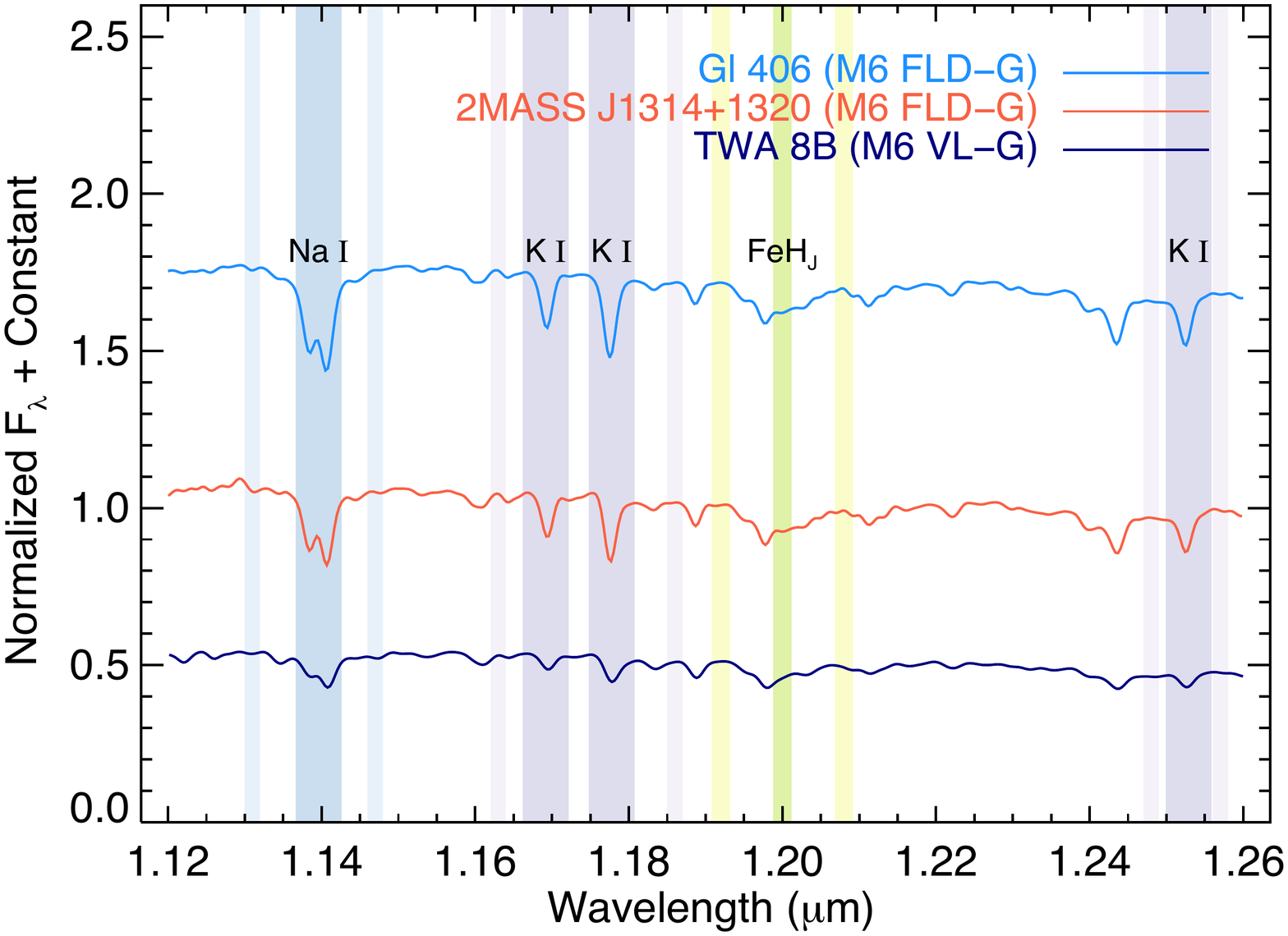}}
  \vskip -0.2in
  \centerline{\includegraphics[height=3.5in]{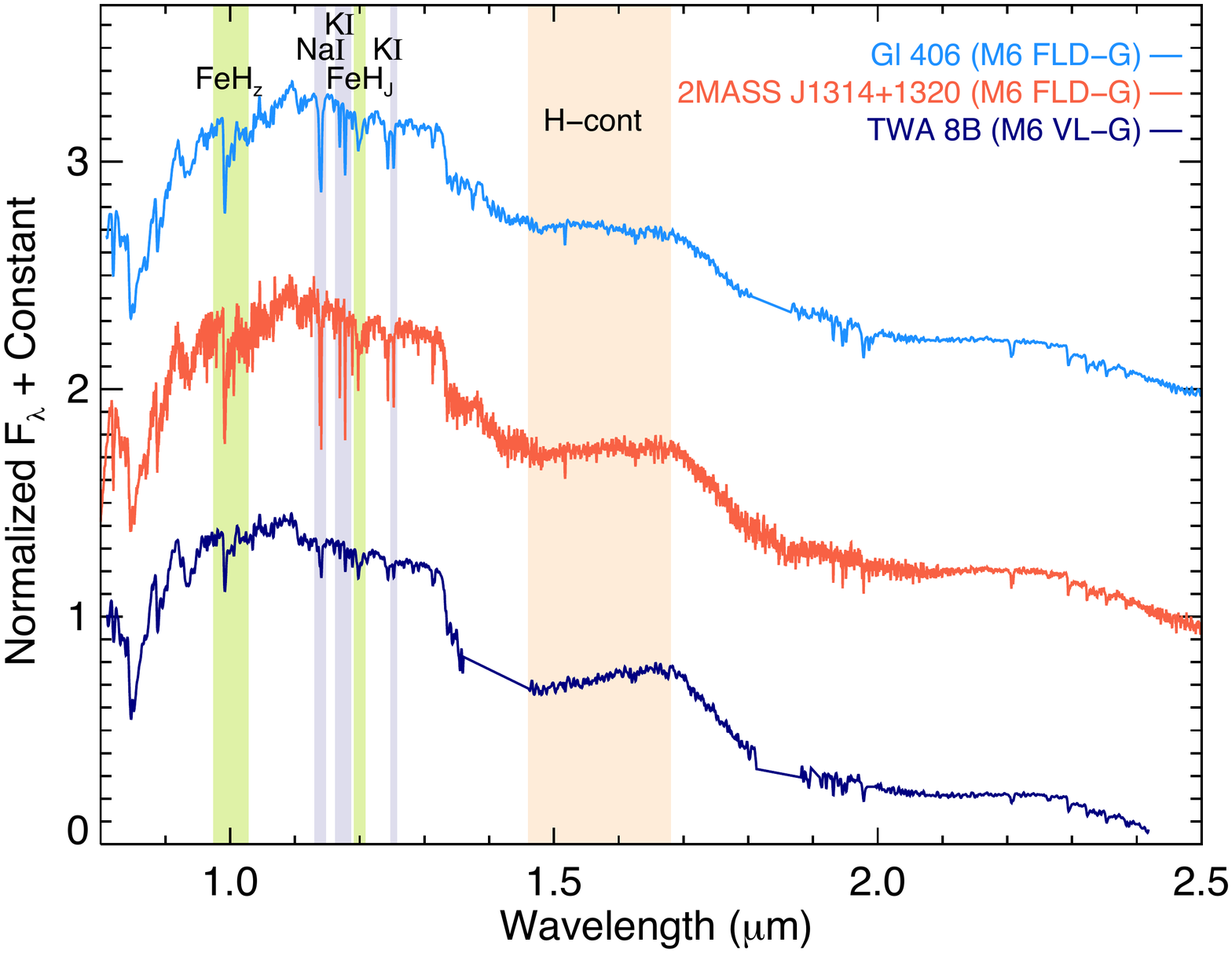}}

  \caption{\normalsize \emph{Top:} a portion of our SpeX spectrum in
    $J$ band shown in comparison to M6 dwarfs with field gravity
    (\fldg) and very low gravity (\vlg).  Although the alkali lines
    visually appear intermediate between field- and low-gravity M6
    dwarfs, a full analysis of all gravity scores from
    \citet{2013ApJ...772...79A} yields a classification of \fldg.
    \obj{AB} is best described as having a spectrum on the borderline
    between \fldg\ and \intg\ classifications. \emph{Bottom:} The full
    SpeX spectrum with other indices (FeH, VO, and $H$-band continuum
    shape) labeled. \label{fig:ir-grav}}

\end{figure}

We also used our SpeX SXD spectrum to determine a NIR spectral type
and gravity classification on the \citet{2013ApJ...772...79A} system.
We find a spectral type of M$6\pm1$ (Figure~\ref{fig:ir-spt}),
consistent with our optical type and the previous determinations of
M$7.0\pm0.5$ in the optical from \citet{2009AJ....137.4109L} and M6.5
in the NIR from \citet{2014ApJ...783...27S}. The gravity
classification is determined from the amount of deviation in certain
spectral features from field objects of similar spectral type.  If
most features are consistent with the field (i.e., having a score of
zero), then the classification is \fldg.  If most features are deviant
from the field indicating low gravity, then the classification is
\vlg.  Intermediate cases are given the classification \intg.  For
\obj, only 3 scores are used to determine the gravity, since the
strength of VO absorption is not applicable at this spectral type.
Based on the indices computed from our SXD spectrum
(Table~\ref{tbl:grav}), we found that FeH was consisent with the
field, most individual alkali lines were not strongly deviant from the
field, and the $H$-band continuum was marginally inconsistent with the
field.  This results in a score of 0n01 and a formal classification of
\fldg.
However, a closer examination of the alkali lines leads to a more
nuanced interpretation.  Of the four features used in the alkali
score, the \ion{K}{1} line at 1.169\,\micron\ was consistently
indicative of low gravity.  Figure~\ref{fig:ir-grav} shows all four
alkali lines, and in fact all of them appear visually weaker than the
field gravity M6 standard, consistent with the findings of
\citet{2014ApJ...783...27S} from an independent spectrum.  The other
three are simply not weak enough to qualify as low gravity on the
\citet{2013ApJ...772...79A} system.  Therefore, while the formal
classification alone gives \fldg, we note that \obj\ is more
faithfully described as having a spectrum on the borderline between
\fldg\ and \intg\ classifications.

Finally, we also used the $K$-band portion of our SpeX spectrum to
determine the metallicity of \obj{AB} in integrated light.  Using the
calibration of \citet{2014AJ....147..160M} we found ${\rm [Fe/H]} =
0.04\pm0.08$\,dex, i.e., consistent with solar metallicity as assumed
by theoretical models in the following analysis.  We caution that this
calibration was based on field objects and so might give a somewhat
different value for [Fe/H] than a relation based on low gravity
dwarfs.

\subsection{Evolutionary Model-Derived Properties \label{sec:evol}}

Given our precisely determined individual masses and luminosities for
the components of \obj{AB}, we can uniquely infer other physical
properties from evolutionary model tracks.  We first interpolate the
values of each physical parameter such as \Teff\ from the BHAC15
evolutionary tracks \citep{2015A&A...577A..42B} onto uniform, 2-d
grids in log(mass) and log(\Lbol), using grid steps of 0.01\,dex in
both axes.  We draw random, normally distributed values of the
individual component's bolometric fluxes for each step in our MCMC
chains.  We then bilinearly interpolate each resultant pair of (mass,
\Lbol) from our chain on a given 2-d grid of parameter values to
compute the posterior distributions of that parameter.  This approach
preserves covariances between input parameters, e.g., mass and \Lbol\
both depend on distance, when deriving parameters like \Teff\
\citep[e.g.,][]{2008ApJ...689..436L}.

The resulting posterior distributions for model-derived values of age,
\Teff, radius, \logg, and fraction of lithium remaining (Li/Li$_{\rm
  init}$) are summarized in Table~\ref{tbl:props}.  We find
model-derived ages that are consistent with coevality at 0.7$\sigma$,
giving a consensus age of $80.8\pm2.5$\,Myr.  (This and other mean
values given in Table~\ref{tbl:props} represent the posterior
distribution of the mean of primary and secondary values calculated
from each step of the chain.)  Other model-derived parameters are
comparably consistent between the two components, as expected given
the very similar component masses and luminosities, with mean values
of $\Teff = 2950\pm4$\,K, $\logg = 4.839\pm0.009$\,dex, $R =
1.820\pm0.016$\,\Rjup, and Li/Li$_{\rm init} = 0.15^{+0.05}_{-0.06}$.
The very small formal uncertainties in our model-derived properties
reflect the precision of the measured masses and luminosities
projected onto the model grids; we do not attempt to include any
systematic errors that could be associated with the models.

Evolutionary models indicate that the components of \obj{AB} are well
removed from the main sequence.  For our measured masses of
$0.0885\pm0.0006$\,\Msun\ and $0.0875\pm0.0010$\,\Msun, BHAC15 models
predict a main sequence luminosity of $\log(\Lbol/\Lsun) = -3.25$\,dex
attained by an age of $\approx$900\,Myr.  Our measured luminosities,
$\log(\Lbol/\Lsun) = -2.617\pm0.010$\,dex and $-2.629\pm0.010$\,dex,
are a factor of $\approx$4$\times$ higher than the main sequence
value.  The detection of lithium by \citet{2014ApJ...783...27S} in the
combined light spectrum of these $\approx$0.09\,\Msun\ objects further
supports the pre--main-sequence nature of this binary, as models
predict lithium will be destroyed in objects of this mass hundreds of
Myr before they reach the main sequence.


\section{Discussion}

\subsection{Lithium Depletion \& Age \label{sec:lith}}

For very low-mass stars ($\lesssim$0.1\,\Msun) and high mass brown
dwarfs, lithium is destroyed at a slow enough rate that it can be used
to determine the ages of stellar associations up to at least
$\sim$100\,Myr.  Higher mass objects destroy their primordial lithium
at a faster rate, so at older ages lithium disappears from the spectra
of progressively lower mass objects.  Therefore, the brightest stars
to display \ion{Li}{1} (6708\,\AA) absorption in a given cluster
define an empirical boundary that can be used as a relative age scale
between different clusters, and evolutionary models can be used to
infer absolute ages of individual clusters based on the location of
this lithium depletion boundary \citep[e.g.,][]{1997ApJ...482..442B,
  2014MNRAS.438L..11B, 2014AJ....147..146K}.  \obj{AB} allows us to
test these model predictions with stars of known mass for the first
time at ages comparable to that of nearby open clusters.

\citet{2014ApJ...783...27S} measured a \ion{Li}{1} (6708\,\AA)
pseudo-equivalent width of ${\rm EW} = 0.46$\,\AA\ from the
integrated-light spectrum of \obj{AB}.  They noted that this is
consistent with measurements of comparable Pleiades late-M dwarfs,
e.g., \citet{1998ApJ...499L.199S} report three M7 dwarfs that all have
${\rm EW} = 0.5$--0.6\,\AA.  Given our mass and luminosity
measurements for each component, BHAC15 evolutionary models predict
that the fraction of initial lithium remaining in the primary and
secondary components is $0.12_{-0.05}^{+0.03}$ and $0.17\pm0.07$,
respectively.  Since the component masses are nearly equal, the
predicted lithium depletion is correspondingly consistent within
1.2$\sigma$.  The mean lithium fraction of the two components is
predicted to be Li/Li$_{\rm init} = 0.15_{-0.06}^{+0.05}$.

An equivalent width measurement cannot be directly converted into
lithium abundance. \citet{2007ApJ...659L..41P} used a model dependent
curve of growth approach to estimate the relationship between
\ion{Li}{1} pseudo-equivalent widths and lithium abundance, defined by
$A({\rm Li}) \equiv \log(N({\rm Li})/N({\rm H})) + 12$.  If the
lithium depletion level in the components of \obj{AB} is $\approx$0.15
as predicted by evolutionary models, and the initial cosmic lithium
abundance is $A({\rm Li}) \approx 3.3$\,dex
\citep{1989GeCoA..53..197A}, then their present day abundance would be
$A({\rm Li}) \approx 2.5$\,dex.  Over a range of $\Teff =
3100$--3600\,K at $\logg = 4.5$\,dex, \citet{2007ApJ...659L..41P}
found that $A({\rm Li}) = 2.5$\,dex corresponds to EW =
0.41--0.51\,\AA, which is consistent with the lithium detection from
\citet{2014ApJ...783...27S}.  If this calibration from
\citet{2007ApJ...659L..41P} is accurate and applicable at the somewhat
lower \Teff\ here ($\approx$3000\,K according to evolutionary models;
Section~\ref{sec:evol}), then the detection of lithium absorption is
fully consistent with the evolutionary model prediction that the
components of \obj{AB} have depleted most of their initial lithium
supply.

As discussed in Section~\ref{sec:evol}, models predict an age of
$80.8\pm2.5$\,Myr based on the mass and luminosity of the components
of \obj{AB}.  We can place this in the context of the relative age
scale provided by nearby open clusters with well determined lithium
depletion boundaries.  For the Pleiades, the boundary is at $M_{K_S} =
8.78\pm0.05$\,mag \citep{2015ApJ...813..108D} using the VLBI parallax
distance of \citet{2014Sci...345.1029M}.  The components of \obj{AB}
are 0.4--0.5\,mag brighter than this (Table~\ref{tbl:props}), implying
that they must be significantly younger than the Pleiades in order to
still possess lithium.  Likewise for Blanco~1,
\citet{2010ApJ...725L.111C} found a lithium depletion boundary of
$\Mbol = 11.99\pm0.30$\,mag, $\approx$0.7\,mag fainter than the
components of \obj{AB}, so they must also be younger than Blanco~1.
The younger cluster $\alpha$~Persei has a lithium depletion boundary
of $\Mbol = 11.31$\,mag \citep{2004ApJ...614..386B}, which is actually
consistent within the errors for both components of \obj{AB}.
Therefore, we conclude that the age of \obj{AB} must be consistent
with or younger than that of $\alpha$~Per. \citet{2004ApJ...614..386B}
report an age of $85\pm10$\,Myr for $\alpha$~Per, and recent age
determinations for the Pleiades and Blanco~1 are, respectively,
$112\pm5$\,Myr \citep{2015ApJ...813..108D} and $132\pm24$\,Myr
\citep{2010ApJ...725L.111C}.  The age of \obj{AB} that we derived from
models using mass and luminosity ($80.8\pm2.5$\,Myr) is therefore
consistent with the requirement from lithium for the system age to be
equal to or younger than the age of $\alpha$~Per.

Finally, we compare the integrated-light color and resolved absolute
magnitudes of \obj{AB} to the cluster sequences of $\alpha$~Per and
the Pleiades. Using our observed spectrum, we compute integrated-light
apparent magnitudes on the Cousins system ($I_C = 11.87\pm0.05$\,mag)
and SDSS system ($i^{\prime} = 12.83\pm0.04$\,mag).  These give
integrated-light colors of $I_C-K_S = 3.08\pm0.05$\,mag and
$i^{\prime}-K_S = 4.04\pm0.04$\,mag.  The $\alpha$~Per members from
\citet{2002A&A...395..813B} within 0.15\,mag of this color have
apparent magnitudes of $K_S = 14.1\pm0.3$\,mag, and assuming a
distance of $172.4\pm2.7$\,pc \citep{2009A&A...497..209V} gives an
absolute magnitude of $M_{K_S} = 7.9\pm0.3$\,mag.  This is somewhat
brighter than but consistent with the absolute magnitudes of the
\obj{AB} components ($M_{K_S} = 8.32\pm0.02$\,mag and
$8.40\pm0.02$\,mag).  Performing the same exercise for the DANCe
sample of probable ($p>0.99$) Pleiades members from
\citet{2015A&A...577A.148B} using our $i^{\prime}-K_S$ color gives
$K_S = 14.2\pm0.3$\,mag and thereby $M_{K_S} = 8.5\pm0.3$\,mag
assuming a Pleiades distance of $136.2\pm1.2$\,pc
\citep{2014Sci...345.1029M}.  This is somewhat fainter than but
consistent with the absolute magnitudes of the \obj{AB} components.
The scatter in these cluster sequences on the color--magnitude diagram
($\approx$0.3\,mag) is relatively large compared to the change of
$\approx$0.6\,mag in absolute magnitude from $\alpha$~Per to the
Pleiades, which limits the discriminating power of this comparison.
The components are somewhat fainter than expected for being as old or
younger than $\alpha$~Per (0.4--0.5\,mag), but the effect is not
significant.  We therefore conclude that the location of the
components of \obj{AB} on the color--magnitude diagram is consistent
with the more precise constraints on age from the lithium depletion
boundary comparison, being comparable age to $\alpha$~Per and younger
than the Pleiades.

\subsection{Membership Assessment  \label{sec:mem}}

Combining our proper motion and parallax with the published radial
velocity of the \obj{AB} system allows us to derive its space motion
and thereby assess potential membership in known associations of young
stars.  \citet{2014ApJ...783...27S} reported spectrally resolved
radial velocities of the two components of \obj{AB} from which they
computed a system velocity of $-10.4\pm1.0$\,\kms\ under the
assumption that the two components are equal in mass.  Our astrometric
mass ratio of $q = 0.989\pm0.007$ now validates this assumption within
their measurement uncertainty and thereby their reported system
velocity.  

\begin{figure} 

  \centerline{\includegraphics[width=6.5in]{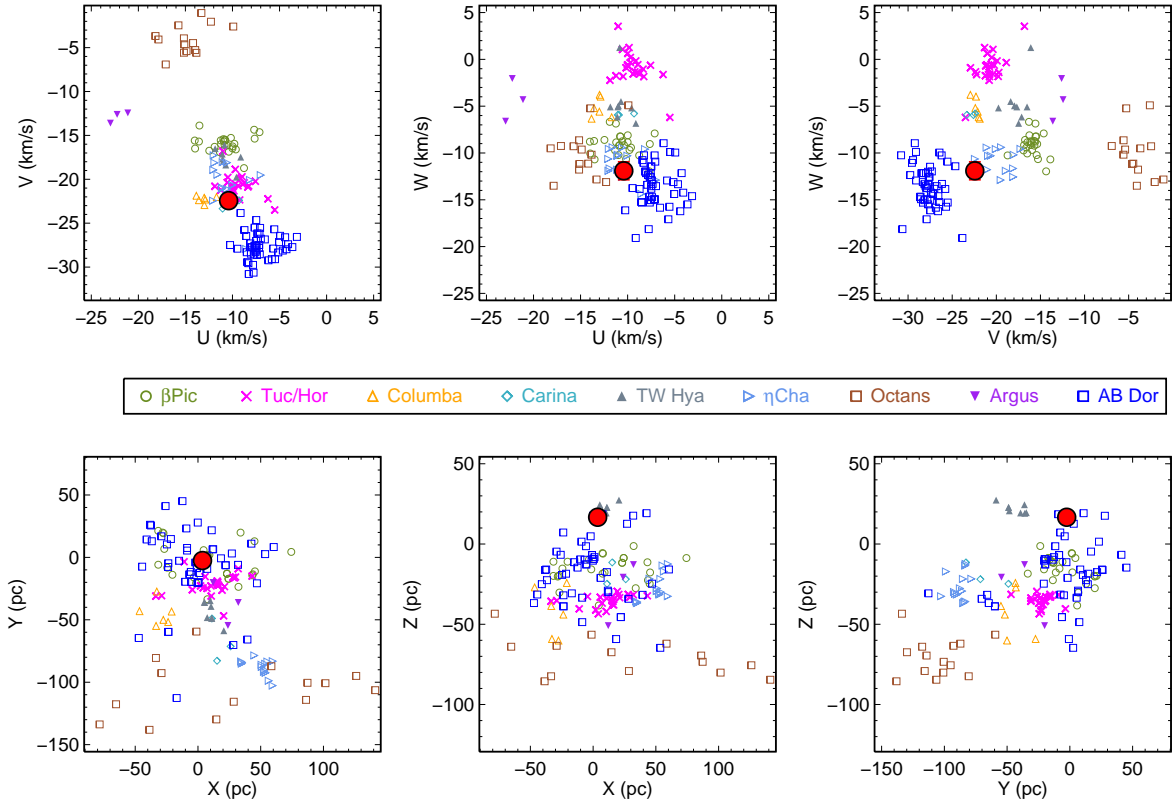}}

  \caption{The heliocentric space velocities and positions for the
    \obj\ system (large red circle; error bars are smaller than the
    symbol size) compared to various young associations (small colored
    symbols).  For these groups we use the known members from
    \citet{2008hsf2.book..757T} that have membership probabilities of
    at least 75\% and parallaxes. We have used RVs and parallaxes from
    the literature for objects which had no measured values in
    \citet{2008hsf2.book..757T}.  The \obj\ system does not seem to be
    associated with any known young moving groups.  \label{fig:uvw}}

\end{figure}

We derive a space motion of $(U, V, W) = (-10.4\pm0.21, -22.27\pm0.15,
-11.9\pm1.0)$\,\kms\ and plot this vector alongside various known
young associations in Figure~\ref{fig:uvw}.  There are no clear visual
associations, except perhaps with $\eta$~Cha in $UVW$, but \obj\ is
very far from $\eta$~Cha in $XYZ$.
Using the BANYAN~II web tool \citep[v1.4;][]{2013ApJ...762...88M,
  2014ApJ...783..121G}, we find a 99.98\% membership probability in
the young field population assuming the age of the system is
$<$1\,Gyr.  Therefore, according to BANYAN, \obj\ is not likely a
member of any of the seven young moving groups considered in their
model (AB~Dor, Argus, $\beta$~Pic, Carina, Columba, Tuc-Hor, and TWA).

We therefore conclude that the \obj{AB} does not belong to any known
young association, despite being unambiguously pre--main-sequence and
located at only 17.25\,pc.  It is possible that \obj{AB} belongs to an
as yet unidentified $<$100~Myr association in the solar neighborhood,
and if so this should be testable with the upcoming release of \Gaia\
astrometry \citep{2012Ap&SS.341...31D}.  If no new associations are
found that match \obj{AB}, then it will join the growing ranks of
orphaned young objects in the solar neighborhood (e.g., Liu et al.\
2016, submitted).

\subsection{Radius \& Effective Temperature \label{sec:teff}}

In principle, the rotation period ($P_*$) and $v\sin(i_*)$
measurements from \citet{2011ApJ...741...27M} and
\citet{2015ApJ...799..192W} provide an empirical constraint on the
minimum stellar radius, $R\sin(i_*)$.\footnote{We use the notation
  $P_*$ and $i_*$ here to indicate to the stellar rotation period and
  the inclination of the stellar rotation axis with respect to the
  plane of the sky, respectively, since we have already used $P$ and
  $i$ for the binary orbit's period and inclination.}  In practice,
such a calculation is complicated by the fact that their measurements
are made in integrated light, but for the sake of argument we will
assume that they both correspond to one of the two components of
\obj{AB}.  \citet{2011ApJ...741...27M} reported a rotation period of
$3.89\pm0.05$\,hr from multi-epoch VLA observations and a $v\sin(i_*)
= 45\pm5$\,\kms\ from optical spectroscopy. In optical photometric
monitoring, \citet{2015ApJ...799..192W} find two distinct rotation
periods of $3.7859\pm0.0001$\,hr and $3.7130\pm0.0002$\,hr and
conclude that the cause of the two very similar but distinct periods
is not clear. Assuming $45\pm5$\,\kms\ and 3.8\,hr gives $R\sin(i_*) =
1.37\pm0.15$\,\Rjup. This is consistent with the model-derived average
$R = 1.820\pm0.016$\,\Rjup, which would correspond to $i_* =
49\pm8\degree$. Interestingly, this value is in good agreement with
the measured orbital inclination $i = 49.34^{+0.28}_{-0.23}$\degree,
which would be consistent with a stellar spin axis aligned with the
orbital plane.  However, it is also possible that the true radius is
smaller or larger than predicted by models, corresponding to $i_* >
49$\degree\ or $i_* < 49$\degree, respectively.

\begin{figure} 

  \centerline{\includegraphics[width=5.0in]{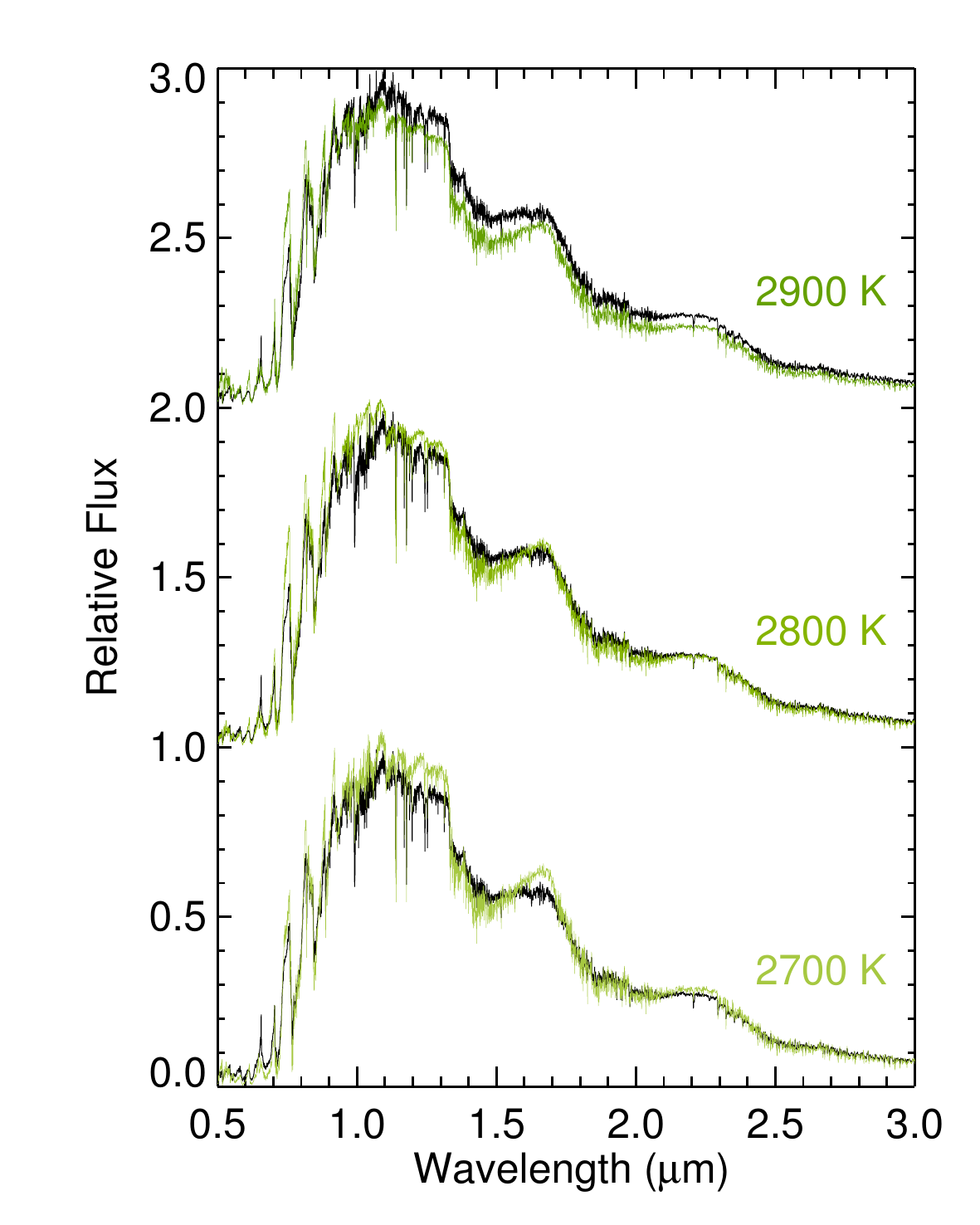}}

  \caption{Our combined SNIFS+SpeX integrated-light spectrum of
    \obj{AB} shown in black compared to BT-Settl model atmosphere
    spectra that all have surface gravity $\logg = 4.5$\,dex shown in
    shades of green.  We fitted a full grid of BT-Settl model spectra
    to our data, and the best fit spectrum had $\Teff = 2800$\,K and
    $\logg = 4.5$\,dex. \label{fig:btsettl}}

\end{figure}

The BHAC15 evolutionary models that we have used to derive stellar
parameters like radius and \Teff\ employ BT-Settl model atmospheres as
their boundary conditions for the surfaces of stars. Therefore, we can
test for consistency between the value of \Teff\ derived from BHAC15
and an independent estimate based on the spectrum of \obj{AB}.
Direct fitting of our combined optical and NIR spectrum with BT-Settl
models yields $\Teff = 2800$\,K and $\logg = 4.5$\,dex (Figure
\ref{fig:btsettl}), but these values necessarily lack in precision due
to the somewhat coarse (100\,K, 0.5\,dex) model grid steps.
The spectral type--\Teff\ scale of \citet{2014ApJ...786...97H}, which
is based on BT-Settl model atmospheres, gives $\Teff = 2770$\,K for a
spectral type of M7, in good agreement with the direct fitting.
Therefore, we find the BT-Settl models give $\approx$180\,K cooler
values for \Teff\ than the average evolutionary model derived value of
$2950\pm4$\,K.\footnote{The spectral type--\Teff\ scale of
  \citet{2003ApJ...593.1093L} gives a somewhat hotter $\Teff =
  2880$\,K at M7, in better agreement with the BHAC15 \Teff. However,
  this could be due to the fact that the \citet{2003ApJ...593.1093L}
  scale was designed to match evolutionary model isochrones of an
  earlier generation of the BHAC15 models and thus does not provide a
  truly independent \Teff\ as needed for our consistency check.
  Moreover, using a spectral type--\Teff\ scale intended for higher
  gravity field dwarfs \citep[e.g.,][]{2013A&A...556A..15R} results in
  an even cooler, more inconsistent \Teff.}

In the absence of additional information, it is equally possible that
this \Teff\ discrepancy could be caused by systematic errors in either
or both of the evolutionary and atmosphere models.  Indeed, the
BT-Settl models do not fit the overall spectrum with high accuracy
(Figure~\ref{fig:btsettl}), implying that any temperature based on
these models will harbor some systematic error.
Observations of other pre--main-sequence binaries at younger ages show
discrepancies consistent with the 180\,K difference here.  For
example, \citet{2016ApJ...817..164R} found that for two unequal-mass
M~dwarf binary systems in the 10-Myr-old Upper Scorpius subgroup, the
evolutionary model-derived temperatures were 100--300\,K higher than
model atmospheres.  If the \Teff\ discrepancy for \obj{AB} were due to
evolutionary models, then at fixed luminosity this would imply model
radii that are too small by 13\%.  Interestingly, there are other
cases of pre--main-sequence M~dwarfs for which such under-predicted
model radii can explain observed discrepancies.
\citet{2015ApJ...807....3K} found that for the 10\,Myr old M5
eclipsing binary UScoCTIO~5, multiple evolutionary models (including
BHAC15) underpredicted the component radii by 10\%--15\% while
simultaneously overpredicting \Teff\ by $\approx$300\,K and thereby
predicting consistent luminosities.
In addition, evolutionary models of low-mass stars that include the
effects of magnetic fields predict larger radii due a slowing down of
their contraction \citep[e.g.,][]{2010ApJ...723.1599M,
  2014ApJ...792...37M, 2016IAUS..314...79F, 2016arXiv160408036F}.

Therefore, both observations and theory suggest that the dominant
source of the \Teff\ discrepancy we observe for \obj{AB} is most
likely due to evolutionary model radii, although we note that spectral
type--\Teff\ scale could still harbor systematic errors.
\obj{AB} is much older than other pre--main-sequence systems with
dynamical mass measurements in nearby star-forming regions, implying
that the same qualitative radius/\Teff\ problem with evolutionary
models extends to ages of at least $\sim$80\,Myr.

\subsection{Spectral Signatures of Low Gravity  \label{sec:grav}}

To our knowledge, this is the first mass measurement for an ultracool
dwarf (spectral type $\gtrsim$M7) with spectral signatures of lower
surface gravity than typical field objects. In Section~\ref{sec:spt},
we classified the integrated-light spectrum as \fldg, but on the
borderline of being \intg, and both \citet{2014ApJ...783...27S} and we
note the presence of spectral features indicative of lower surface
gravity relative to field dwarfs of similar spectral type.
Evolutionary models indicate that the surface gravities of the binary
components are $\logg = 4.83$--4.87\,dex, which is 0.42--0.46\,dex
lower than the predicted main sequence surface gravity of 5.29\,dex
for a 0.09\,\Msun\ star. If the model radii are too small by 13\% as
suggested above, then the model-derived gravities would be 0.1\,dex
lower. Overall, this implies that the borderline between \fldg\ and
\intg\ designations for late-M dwarfs corresponds to a surface gravity
$\approx$0.5\,dex lower than field objects.

While the \obj{AB} system itself does not belong to an association of
independently determined age, numerous other ultracool dwarfs with
gravity classifications do.  This allows us to check for consistency
of spectral behavior between \obj{AB} and other young late-M dwarfs of
known age. In the AB~Dor moving group (125\,Myr),
\citet{2016ApJ...821..120A} list 15 bona fide or strong candidate
members. Of these, all but one object have gravity classifications of
\intg\ or \vlg, with only 2MASS~J03264225$-$2102057 (L5) classified as
\fldg\ (but possessing some visual signs of youth, like \obj{AB} but
at much later type). Of 17 possible candidate members none are
classified \fldg.
Moreover, \citet{2013MmSAI..84.1089A} report gravity classifications
for eight ultracool dwarfs in the Pleiades; none were \fldg, and the
two M7 dwarfs PPl~1 and Teide~1 were \intg\ and \vlg, respectively.
These comparisons imply that spectral signatures leading to \intg\
and \vlg\ classifications are typically still quite entrenched at ages
significantly older than \obj{AB}. At even younger ages, e.g., Tuc-Hor
(50\,Myr) and $\beta$~Pic (23\,Myr), no known members are classified
as \fldg\ either (e.g., see compilation of Liu et al.\ 2016,
submitted). Therefore, the fact that \obj{AB} has a gravity
classification of \fldg\ is at odds with its model-derived age of
$80.8\pm2.5$\,Myr and that our lithium analysis that empirically
places the system at a significantly younger age than the Pleiades.

We suggest a few possible explanations for the fact that \obj{AB}
shows less distinct evidence for low gravity than older ultracool
dwarfs.  Perhaps it is not that uncommon for late-M dwarfs to have
more muted gravity signatures than L~dwarfs, and the existing samples
of ultracool dwarfs are too sparse to detect this yet.  As noted by
\citet{2013MmSAI..84.1089A}, the currently available infrared spectra
of Pleiades late-M dwarfs are of much lower S/N than was used to
define the gravity classification system, so higher quality spectra
may result in somewhat different, higher gravity, classifications.
The classification of \obj{AB} is done in integrated light, so perhaps
the gravity signatures are somehow obscured in combined light.
\citet{2013MmSAI..84.1089A} tested such an idea, mostly using \vlg\
templates for components, and found that it is quite rare for two
components to be classified as lower gravity in combined light.
Finally, maybe some third parameter is at work, e.g., metallicity or
rotation, causing the gravity classification to deviate slightly from
other young ultracool dwarfs. Our spectrum shows no signs of unusual
metallicity, and it also seems unlikely that such a young object would
different substantially in composition from other young associations
in the solar neighborhood \citep[e.g.,][]{2008A&A...480..889S}.

\subsection{H-R Diagram Test \label{sec:hrd}}

\begin{figure} 

  \centerline{\includegraphics[width=5.0in]{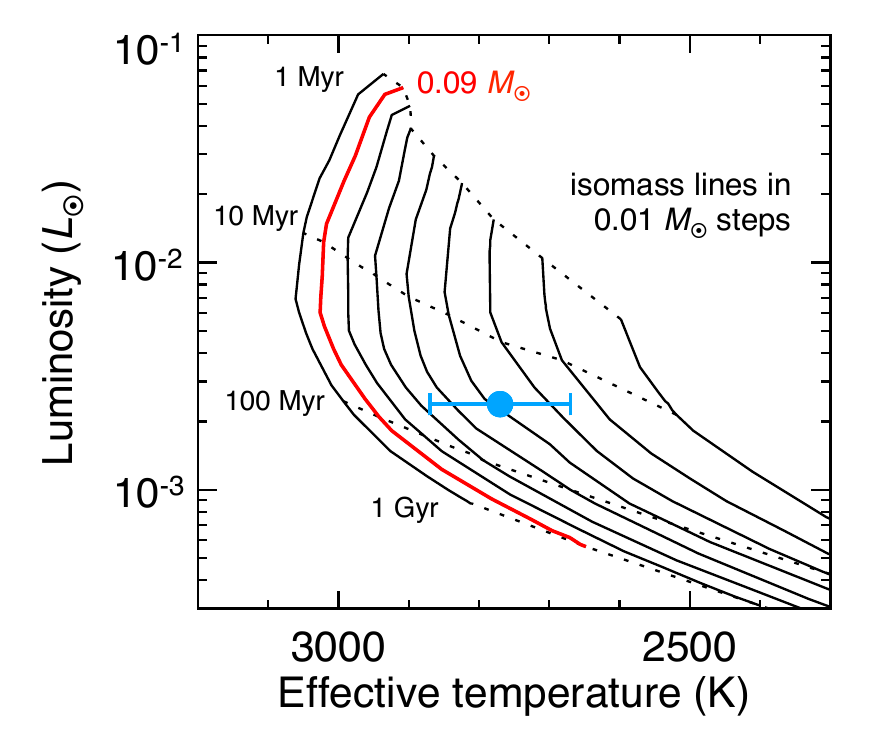}}

  \caption{H-R diagram showing the mean luminosity of the \obj{AB}
    components and the \Teff\ determined from the integrated-light
    spectral type of M7.0 (blue data point) compared to evolutionary
    model tracks.  BHAC15 isomass tracks are shown in steps of
    0.01\,\Msun\ with the 0.09\,\Msun\ track highlighted in red since
    this is consistent within $<$3\% of both measured individual
    masses.  Isochrones from 1\,Myr to 1\,Gyr are indicated by dotted
    lines.  The \Teff\ used for \obj{AB} here is calibrated off of the
    BT-Settl model atmospheres, the same used for boundary conditions
    in the BHAC15 evolutionary models.  When we use this \Teff\ and
    luminosity to infer properties from BHAC15 models, we find a mass
    $46^{+16}_{-19}$\% (2.0$\sigma$) lower than we measured
    dynamically.  This discrepancy indicates either large errors in
    spectral type--\Teff\ relations ($\approx$180\,K) or systematic
    errors in evolutionary models (e.g., 13\% in radius).  In either
    case, this result suggests that masses inferred for young stars
    from the H-R diagram will harbor large systematic errors, and
    young stars may be confused for young brown
    dwarfs.  \label{fig:hrd}}

\end{figure}

Pre--main-sequence stellar models are commonly used to infer masses by
placing objects on the H-R diagram
\citep[e.g.,][]{1998ApJ...508..347L, 2002AJ....124..404P,
  2012ApJ...748...14D}.  To test the accuracy of masses derived from
models in this way, we used the effective temperatures and
luminosities of \obj{AB} to derive mass and age.  Given that the
masses and luminosities of the components are nearly equal, we simply
consider the average integrated-light properties for this test. As
mentioned above, the spectral type--\Teff\ scale for young objects
from \citet{2014ApJ...786...97H} gives $\Teff = 2770$\,K, and they
estimate a systematic error of 100\,K for their \Teff\ scale, which we
adopt here. The mean measured luminosity of \obj{AB} is
$\log(\Lbol/\Lsun) = -2.623\pm0.010$\,dex.  Figure~\ref{fig:hrd} shows
these values of \Teff\ and \Lbol\ compared to BHAC15 evolutionary
model tracks.

For this test, we interpolated model tracks on a uniform grid of
$\log(\Lbol)$ and $\log(\Teff)$ in the same fashion as described above
in Section~\ref{sec:evol}. We found an H-R diagram derived average
component mass of $50^{+13}_{-20}$\,\Mjup, age of
$25^{+10}_{-17}$\,Myr, and \logg\ of $4.46^{+0.20}_{-0.22}$\,dex. This
would imply that \obj{AB} is actually a pair of young brown dwarfs,
due to the H-R diagram derived mass being $46^{+16}_{-19}$\%
(2.0$\sigma$) smaller than our directly measured component masses of
$\approx$92\,\Mjup. The H-R diagram age is also much smaller
($0.5\pm0.3$\,dex) than the age derived from the same models using
mass and \Lbol. As expected, the mass and age posteriors derived from
the H-R diagram are highly correlated, where lower masses correspond
to younger ages.

The discrepancy between the H-R diagram derived mass and our
dynamically measured masses suggests either large errors in the
spectral type--\Teff\ relations, which are calibrated using BT-Settl
model atmospheres, systematic errors in the evolutionary models, or
some combination of both things.  As we discuss in
Section~\ref{sec:teff}, we suggest the dominant source of this
discrepancy is that evolutionary model radii are underpredicted and
\Teff\ is thereby overpredicted at a given luminosity.
Regardless of the cause of the discrepancy, this test case shows that
masses derived from the H-R diagram can harbor large systematic
errors. Unfortunately, this method is often the only practical option
when attempting to infer masses of stars and brown dwarfs in young
associations where age can be uncertain due to potential underlying
age spreads.  Therefore, the systematic error we have identified here
will have significant implications for efforts to determine the
low-mass end of the initial mass function, suggesting that young
low-mass stars may be mistakenly identified as young brown dwarfs.


\section{Conclusions}

We present here individual dynamical masses for the components of
\obj{AB}, a pre--main-sequence binary located at a distance of only
$17.249\pm0.013$\,pc. These masses and parallactic distance are made
possible by a joint analysis of our resolved relative astrometry of
the primary and secondary from Keck AO imaging and masking along with
absolute astrometry from VLBA radio interferometry of the secondary
(i.e., the component that is fainter in the optical and infrared). We
also derive component luminosities using integrated-light spectroscopy
and photometry and our resolved infrared photometry. The measured
component masses of $92.8\pm0.6$\,\Mjup\ ($0.0885\pm0.0006$\,\Msun)
and $91.7\pm1.0$\,\Mjup\ ($0.0875\pm0.0010$\Msun) and luminosities of
$\log(\Lbol/\Lsun) = -2.616\pm0.010$\,dex and $-2.631\pm0.010$\,dex,
respectively, are consistent with being coeval at an age of
$80.8\pm2.5$\,Myr according to BHAC15 evolutionary models.
Our precise masses and luminosities are largely thanks to a remarkably
precise VLBA parallax ($\sigma_{\pi}/\pi = 9\times10^{-4}$; Paper~I).
We determine that \obj{AB} is unambiguously in the pre--main-sequence
phase of its evolution based on having lithium absorption and
luminosities $\approx$4$\times$ higher than predicted for the main
sequence at our measured masses.
This combination of precise distance, masses, luminosities, and the
detection of lithium by \citet{2014ApJ...783...27S} enables novel
tests of pre--main-sequence models distinct from previous work on
objects in star forming regions with more uncertain distances.

\begin{enumerate}

\item Evolutionary models self-consistently predict luminosity and
  lithium depletion in this binary thereby passing the first test of
  the theoretical lithium depletion boundary using ultracool dwarfs of
  known mass.  Models predict the components have lost
  $88^{+5}_{-3}$\% and $83\pm7$\% of their initial lithium, leaving
  enough remaining that they are still expected to display lithium
  absorption.  On the empirically defined relative cluster age scale,
  the presence of lithium and the component absolute magnitudes of
  \obj{AB} imply an age consistent with or younger than
  $\alpha$~Persei and significantly younger than the Pleiades and
  Blanco~1.

\item We compare the effective temperature derived from evolutionary
  models given our mass and luminosity (component average $\Teff =
  2950\pm5$\,K) to that derived from spectral type--\Teff\ relations
  based on BT-Settl models ($2770\pm100$\,K). The 180\,K discrepancy
  in these values would correspond to a 13\% radius error at fixed
  luminosity, which is similar to a problem observed for low mass
  stars in the much younger Upper Scorpius star-forming region. The
  eclipsing binary USco~CTIO~5 shows that model radii are likely the
  dominant source of the \Teff\ problem, while the radius and \Teff\
  errors cancel so that model luminosities are accurate.  We suggest
  that the same physical cause is responsible for the 180\,K
  discrepancy for \obj{AB}, showing that this radius problem can
  extend to much older pre--main-sequence ages than previously
  recognized.

\item In a test that mimics the typical application of evolutionary
  models by observers, we used the \Teff\ derived from spectral type
  and the average component luminosity to infer mass and age from
  evolutionary model tracks on the H-R diagram.  The \Teff--\Lbol\
  derived average component mass of $50^{+13}_{-20}$\,\Mjup\ is much
  lower ($46^{+16}_{-19}$\%, 2.0$\sigma$) than we measure dynamically.
  This highlights the large systematic errors possible when inferring
  masses of low-mass stars and brown dwarfs at young ages and implies
  that some stars may be mistakenly identified as brown dwarfs when
  using the H-R diagram.

\item The integrated-light spectrum of \obj{AB} displays signatures of
  low surface gravity, although we formally classify it as \fldg\ (on
  the borderline of \intg) on the infrared \citet{2013ApJ...772...79A}
  system.  This is the first time dynamical masses have been measured
  for ultracool dwarfs with low-gravity spectral features.  However,
  contrary to expectations, \obj{AB} shows less distinct spectral
  signs of low gravity than ultracool dwarfs at older ages (Pleiades,
  AB~Dor), which we are unable to explain.

\end{enumerate}

\obj{AB} provides a high-precision benchmark for pre--main-sequence
models at a distance $\sim$10$\times$ closer than even the nearest
star-forming regions.  One major unresolved question is why only the
secondary component is radio emitting, as discussed in detail in
Paper~I.  Given that we have shown that the component masses are
within 2\% of each other for this coeval, co-compositional binary
system, one likely explanation for the divergent behavior is a
difference in the angular momentum evolution of the two components.
Projected rotational velocities ($v\sin{i}$) of other very low-mass
binaries have hinted at such differences in angular momentum evolution
\citep[e.g.,][]{2012ApJ...750...79K}, but such measurements have yet
to be obtained for \obj{AB}.  The components of \obj{AB} are now the
nearest, lowest mass pre--main-sequence stars with direct mass
measurements.  In the future, \Gaia\ parallaxes combined with ongoing
ground-based orbit monitoring efforts will make many more such tests
of models possible for more distant binaries in star forming regions.
\Gaia\ data could even help identify previously unrecognized
associations of young stars in the solar neighborhood to which
\obj{AB}, currently not associated with any known group, may belong.


\acknowledgments

This work was supported by a NASA Keck PI Data Award, administered by
the NASA Exoplanet Science Institute.
We thank Will Best for assistance with some Keck/NIRC2 observations.
It is a pleasure to thank Joel Aycock, Carolyn Jordan, Jason McIlroy,
Luca Rizzi, Terry Stickel, Hien Tran,
and the Keck Observatory staff for assistance with our Keck AO
observations.  
We also thank
P.~K.~G.\ Williams, Michael Ireland, Katelyn Allers, Joshua Schlieder, and Mark Reid for useful discussions.
The anonymous referee provided a remarkably rapid and thoughtful review that helped refine our discussion.
James R.\ A.\ Davenport for distributing his IDL implementation
of the cubehelix color scheme \citep{2011BASI...39..289G}.
Our research has employed 
the 2MASS data products; 
NASA's Astrophysical Data System;
data from the Wide-field Infrared Survey Explorer, which is a joint
project of the University of California, Los Angeles, and the Jet
Propulsion Laboratory/California Institute of Technology, funded by
NASA and curated by the NASA/IPAC Infrared Science Archive;
and the SIMBAD database operated at CDS, Strasbourg, France.
Finally, the authors wish to recognize and acknowledge the very
significant cultural role and reverence that the summit of Maunakea has
always had within the indigenous Hawaiian community.  We are most
fortunate to have the opportunity to conduct observations from this
mountain.

{\it Facilities:} \facility{IRTF}, \facility{Keck:II (NGS AO, NIRC2)},
\facility{UH:2.2m}

\clearpage

\clearpage

\clearpage

\clearpage

\clearpage

\clearpage

\clearpage

\clearpage

\clearpage

\clearpage
\begin{deluxetable}{lccccc}
\tablewidth{0pt}
\tablecaption{Keck/NIRC2 Adaptive Optics Astrometry for \obj{AB} \label{tbl:keck}}
\tablehead{
\colhead{Date (UT)}            &
\colhead{Filter}               &
\colhead{Separation (mas)}     &
\colhead{PA (\degree)}         &
\colhead{$\Delta{m}$ (mag)}    &
\colhead{Note}                 }
\startdata
2013~Jan~18 & $K$ & \phn$  66.2\pm0.5 $ &     $202.3\pm0.5 $ & $0.092\pm0.012$ & masking \\
2014~Jan~22 & $K$ & \phn$  54.0\pm1.9 $ &     $  342\pm7   $ & $0.048\pm0.024$ & masking \\
2014~May~9  & $K$ & \phn$ 78.62\pm0.21$ & \phn$ 15.2\pm0.6 $ & $ 0.11\pm0.04 $ & imaging \\
2014~Jun~15 & $J$ & \phn$  88.1\pm0.5 $ & \phn$ 21.7\pm0.5 $ & $ 0.08\pm0.04 $ & imaging \\
2015~Jan~14 & $K$ &     $137.25\pm0.19$ & \phn$41.55\pm0.04$ & $0.091\pm0.006$ & imaging \\
2015~Apr~10 & $K$ &     $153.16\pm0.08$ & \phn$46.27\pm0.03$ & $0.068\pm0.005$ & imaging \\
2015~Jun~28 & $K$ &     $166.43\pm0.13$ & \phn$49.84\pm0.05$ & $0.081\pm0.006$ & imaging \\
\enddata
\end{deluxetable}

\clearpage
\begin{landscape}
\begin{deluxetable}{lcccc}
\setlength{\tabcolsep}{0.10in}
\tabletypesize{\tiny}
\tablewidth{0pt}
\tablehead{
\colhead{Property}              &
\colhead{Median $\pm$1$\sigma$} &
\colhead{Best fit}              &
\colhead{95.4\% c.i.}           &
\colhead{Prior/Notes}           }
\tablecaption{MCMC Posteriors for the Orbit and Parallax of LSPM~J1314+1320AB \label{tbl:mcmc}}
\startdata
\multicolumn{5}{c}{Fitted parameters} \\[1pt]
\cline{1-5}
\multicolumn{5}{c}{} \\[-5pt]
Orbital period $P$ (yr)                                         & $9.58_{-0.08}^{+0.07}$         & 9.58          &         9.45, 9.74         & $1/P$ (log-flat)                                          \\[3pt]
Semimajor axis $a = a_1 + a_2$ (mas)                            & $146.6\pm0.5$\phn\phn          & 146.4         &        145.6, 147.7        & $1/a$ (log-flat)                                          \\[3pt]
Eccentricity $e$                                                & $0.6011_{-0.0025}^{+0.0022}$   & 0.6014        &       0.5964, 0.6060       & uniform, $0 \leq e < 1$                                   \\[3pt]
Inclination $i$ (\degree)                                       & $49.34_{-0.23}^{+0.28}$\phn    & 49.19         &        48.77, 49.82        & $\sin(i)$, $0\degree < i < 180\degree$                    \\[3pt]
PA of the ascending node $\Omega$ (\degree)                     & $60.4\pm0.4$\phn               & 60.2          &         59.6, 61.3         & uniform                                                   \\[3pt]
Argument of periastron $\omega$ (\degree)                       & $205.6\pm0.7$\phn\phn          & 205.8         &        204.1, 207.1        & uniform                                                   \\[3pt]
Mean longitude at 2455197.5~JD $\lambda_{\rm ref}$ (\degree)    & $71.8_{-1.4}^{+1.3}$\phn       & 71.9          &         69.2, 74.5         & uniform                                                   \\[3pt]
$\alpha_{2010}-{\rm median}(\alpha_{2010})$ (mas)               & $0.0\pm0.4$                    & 0.0           &       $-$0.8, 0.8\phs      & uniform, ${\rm median}(\alpha_{2010})=198.5841023$\degree \\[3pt]
$\delta_{2010}-{\rm median}(\delta_{2010})$ (mas)               & $0.0\pm0.9$                    & 0.0           &       $-$1.7, 1.7\phs      & uniform, ${\rm median}(\delta_{2010})=+13.3330434$\degree \\[3pt]
Proper motion in RA $\mu_{\alpha\cos\delta}$ (\masyr)           & $-247.99\pm0.10$\phn\phn\phs   & $-$248.01     &    $-$248.19, $-$247.78    & uniform                                                   \\[3pt]
Proper motion in Dec $\mu_{\delta}$ (\masyr)                    & $-183.58\pm0.22$\phn\phn\phs   & $-$183.64     &    $-$184.05, $-$183.17    & uniform                                                   \\[3pt]
Parallax $\pi$ (mas)                                            & $57.975\pm0.045$\phn           & 57.988        &       57.856, 58.082       & $1/\pi^2$ (uniform volume density)                        \\[3pt]
Semimajor axis of secondary $a_2$ (mas)                         & $73.7\pm0.3$\phn               & 73.7          &         73.1, 74.4         & uniform                                                   \\[3pt]
RA VLBI error parameter $\log(\sigma_{\alpha}^2)$ [deg$^2$]     & $-15.5\pm0.4$\phn\phs          & $-$15.8       &      $-$16.2, $-$14.7      & uniform                                                   \\[3pt]
Dec VLBI error parameter $\log(\sigma_{\delta}^2)$ [deg$^2$]    & $-14.6\pm0.4$\phn\phs          & $-$14.7       &      $-$15.4, $-$13.7      & uniform                                                   \\[3pt]
\cline{1-5}
\multicolumn{5}{c}{} \\[-5pt]
\multicolumn{5}{c}{Derived properties} \\[1pt]
\cline{1-5}
\multicolumn{5}{c}{} \\[-5pt]
Total mass $\Mtot$ (\Mjup)                                      & $184.5\pm1.6$\phn\phn          & 183.7         &        181.1, 187.6        & \nodata                                                   \\[3pt]
Primary mass $M_1$ (\Mjup)                                      & $92.8\pm0.6$\phn               & 92.5          &         91.5, 94.0         & \nodata                                                   \\[3pt]
Secondary mass $M_2$ (\Mjup)                                    & $91.7\pm1.0$\phn               & 91.2          &         89.6, 93.8         & \nodata                                                   \\[3pt]
Mass ratio $q \equiv M_2/M_1$                                   & $0.989\pm0.007$                & 0.986         &        0.975, 1.002        & \nodata                                                   \\[3pt]
Distance $d$ (pc)                                               & $17.249\pm0.013$\phn           & 17.245        &       17.217, 17.284       & \nodata                                                   \\[3pt]
Semimajor axis $a$ (AU)                                         & $2.528\pm0.009$                & 2.525         &        2.512, 2.548        & \nodata                                                   \\[3pt]
Time of periastron $T_0$ (JD)                                   & $2456498.5_{-1.8}^{+1.7}$      & 2456498.8     &    2456495.0, 2456502.0    & \nodata                                                   \\[3pt]
\enddata
\end{deluxetable}
\clearpage
\end{landscape}

\clearpage
\begin{deluxetable}{lcccc}
\tabletypesize{\footnotesize}
\tablewidth{0pt}
\tablecaption{Properties of \obj{AB} \label{tbl:props}}
\tablehead{
\colhead{Property}          &
\colhead{\obj{A}}           &
\colhead{\obj{B}}           &
\colhead{Integrated\tablenotemark{*}}  &
\colhead{$\Delta$ = B $-$ A}}
\startdata

Spectral type (optical)                    & \nodata                     & \nodata                     & M$7.0\pm0.2$                & \nodata                     \\
Spectral type (near-IR)                    & \nodata                     & \nodata                     & M$6\pm1$ \fldg\             & \nodata                     \\

Spectral type (optical)                        & \nodata                  & \nodata                  & M$7.0\pm0.2$           & \nodata                        \\
Spectral type (near-IR)                        & \nodata                  & \nodata                  & M$6\pm1$ \fldg\        & \nodata                        \\
$J_{\rm MKO}$ (mag)                            & $10.430\pm0.029$\phn     & $10.51\pm0.03$\phn       & $9.717\pm0.022$        & $0.08\pm0.04$                  \\
$H_{\rm MKO}$ (mag)                            & $9.94\pm0.04$            & $9.97\pm0.04$            & $9.20\pm0.03$          & $0.03\pm0.06$                  \\
$K_{\rm MKO}$ (mag)                            & $9.480\pm0.021$          & $9.560\pm0.022$          & $8.767\pm0.018$        & $0.080\pm0.022$                \\
$J_{\rm 2MASS}$ (mag)                          & $10.467\pm0.029$\phn     & $10.55\pm0.03$\phn       & $9.754\pm0.022$        & $0.08\pm0.04$                  \\
$H_{\rm 2MASS}$ (mag)                          & $9.91\pm0.04$            & $9.94\pm0.04$            & $9.18\pm0.03$          & $0.03\pm0.06$                  \\
$K_{S,{\rm 2MASS}}$ (mag)                      & $9.507\pm0.021$          & $9.587\pm0.022$          & $8.794\pm0.018$        & $0.080\pm0.022$                \\
$M_{J, {\rm MKO}}$ (mag)                       & $9.246\pm0.029$          & $9.33\pm0.03$            & $9.717\pm0.022$        & $0.08\pm0.04$                  \\
$M_{H, {\rm MKO}}$ (mag)                       & $8.76\pm0.04$            & $8.79\pm0.04$            & $9.20\pm0.03$          & $0.03\pm0.06$                  \\
$M_{K, {\rm MKO}}$ (mag)                       & $8.296\pm0.021$          & $8.376\pm0.022$          & $8.767\pm0.018$        & $0.080\pm0.022$                \\
$M_{J, {\rm 2MASS}}$ (mag)                     & $9.283\pm0.029$          & $9.36\pm0.03$            & $9.754\pm0.022$        & $0.08\pm0.04$                  \\
$M_{H, {\rm 2MASS}}$ (mag)                     & $8.73\pm0.04$            & $8.76\pm0.04$            & $9.18\pm0.03$          & $0.03\pm0.06$                  \\
$M_{K_S, {\rm 2MASS}}$ (mag)                   & $8.323\pm0.021$          & $8.403\pm0.022$          & $8.794\pm0.018$        & $0.080\pm0.022$                \\
$\log(f_{\rm bol})$ [erg\,cm$^{-2}$\,s$^{-1}$] & $-9.584\pm0.010$\phs     & $-9.599\pm0.010$\phs     & $-9.291\pm0.009$\phs   & $-0.015\pm0.010$\phs           \\
$M_{\rm bol}$ (mag)                            & $11.290\pm0.025$\phn     & $11.328\pm0.025$\phn     & $11.309\pm0.023$\phn   & $-0.038\pm0.025$\phs           \\
$\log(L_{\rm bol}/\Lsun)$ (dex)                & $-2.616\pm0.010$\phs     & $-2.631\pm0.010$\phs     & $-2.322\pm0.009$\phs   & $-0.015\pm0.010$\phs           \\

\cline{1-5}
\multicolumn{5}{c}{} \\
\multicolumn{5}{c}{Derived from BHAC15 Evolutionary Models} \\
\cline{1-5}
Age (Myr)                                      & $79.9_{-2.7}^{+2.5}$\phn & $81.7_{-3.3}^{+2.9}$\phn & $80.8\pm2.5$\phn       & $1.8\pm2.7$                    \\
\Teff\ (K)                                     & $2954\pm3$\phn\phn\phn   & $2947\pm4$\phn\phn\phn   & $2950\pm4$\phn\phn\phn & $-7\pm3$\phs                   \\
Radius (\Rjup)                                 & $1.831\pm0.018$          & $1.808\pm0.018$          & $1.820\pm0.016$        & $-0.023_{-0.017}^{+0.018}$\phs \\
\logg\ [cm\,s$^{-2}$]                          & $4.836\pm0.010$          & $4.842\pm0.011$          & $4.839\pm0.009$        & $0.006\pm0.009$                \\
Li/Li$_{\rm init}$                             & $0.12_{-0.05}^{+0.03}$   & $0.17\pm0.07$            & $0.15_{-0.06}^{+0.05}$ & $0.05\pm0.04$                  \\

\enddata
\tablenotetext{*}{Directly measured properties have their
  integrated-light values given.  For model-derived properties, we
  report the mean of individually derived values for \obj{A} and
  \obj{B}.}
\end{deluxetable}

\clearpage
\begin{deluxetable}{lc}
\tablewidth{0pt}
\tablecaption{Gravity Classification Summary \label{tbl:grav}}
\tablehead{
\colhead{Name}                  &
\colhead{Index/Score}           }
\startdata
FeH$_{z}$                   &  $1.1088\pm0.0012$            \\
FeH$_{J}$                   &  $1.0722\pm0.0019$            \\
VO$_{z}$                    &  $1.0010_{-0.0010}^{+0.0009}$ \\
\ion{K}{1}$_{J}$            &  $1.0482\pm0.0006$            \\
$H$-cont                    &   $0.9905\pm0.0006$           \\
\ion{Na}{1}                 &  $8.22\pm0.10$                \\
\ion{K}{1} (1.169\,\micron) &  $2.10\pm0.10$                \\
\ion{K}{1} (1.177\,\micron) &  $3.61_{-0.09}^{+0.10}$       \\
\ion{K}{1} (1.253\,\micron) &  $2.63\pm0.08$                \\
Alkali Score                &  1000                         \\
Final Score                 &  0n01                         \\
Gravity Classification      &  \fldg                        \\
\enddata
\tablecomments{The parts of the final score correspond respectively to
  lines of FeH (0), VO (n/a), alkali (0), and $H$-band continuum (1).}
\end{deluxetable}

\end{document}